%% file: main.tex
\documentclass[aps,prd,twocolumn,superscriptaddress,nofootinbib,floatfix,10pt]{revtex4-2}
\usepackage[T1]{fontenc}
\usepackage{amsmath,amssymb,booktabs}
\usepackage{graphicx}
\usepackage[stretch=30,shrink=30]{microtype}
\newcommand{\Tr}{\operatorname{Tr}}
\newcommand{\vev}[1]{\langle #1\rangle}
\newcommand{\R}{\mathbb R}
\newcommand{\Om}{\Omega_8}
\newcommand{\rep}[1]{\mathbf{#1}}
\newcommand{\ket}[1]{|#1\rangle}
\newcommand{\bra}[1]{\langle#1|}
\usepackage[x11names]{xcolor}
\usepackage{silence}
\usepackage{dsfont}
\usepackage[colorlinks=true,linkcolor=RoyalBlue4,citecolor=RoyalBlue4,urlcolor=RoyalBlue4]{hyperref}
\newcommand{\lp}{l_{p}}
\makeatletter
\renewcommand\paragraph{\@startsection{paragraph}{4}{\z@}{1.5ex plus .5ex minus .2ex}{-0.6em}{\normalfont\normalsize\bfseries}}
\makeatother
\makeatletter
\def\@bibdataout@aps{\immediate\write\@bibdataout{@CONTROL{apsrev42Control,author="48",editor="1",pages="0",title="0",year="1"}}\if@filesw\immediate\write\@auxout{\string\citation{apsrev42Control}}\fi}
\makeatother
\makeatletter
\newcommand{\SZtocoff}{\let\SZ@addcontentsline\addcontentsline\def\addcontentsline##1##2##3{}}
\newcommand{\SZtocon}{\let\addcontentsline\SZ@addcontentsline}
\newcommand{\SZbibfont}{\def\bibfont{\small\@clubpenalty\clubpenalty\raggedright}}
\makeatother
\hypersetup{pdftitle={Bound States of Two D0-Branes and the Three-Graviton Coupling},pdfauthor={Sergio Sanjurjo, Themistocles Zikopoulos}}

\begin{document}

\title{Bound States of Two D0-Branes and the Three-Graviton Coupling}
\author{Sergio Sanjurjo}
\email{ssanjurjo@perimeterinstitute.ca}
\affiliation{Perimeter Institute for Theoretical Physics, Waterloo, Ontario, N2L 2Y5, Canada}
\affiliation{Department of Physics and Astronomy, University of Waterloo, Waterloo, Ontario, N2L 3G1, Canada}
\author{Themistocles Zikopoulos}
\email{tzikopoulos@g.harvard.edu}
\affiliation{Jefferson Physical Laboratory, Harvard University, Cambridge, MA 02138, USA}

\begin{abstract}
\input{Sections/0.Abstract}
\end{abstract}

\maketitle

\input{Sections/1.Introduction}
\input{Sections/2.BoundState}
\input{Sections/3.Ward}
\input{Sections/4.Excited}
\input{Sections/5.Outlook}

\clearpage
\onecolumngrid
\par\vspace{\baselineskip}
\textit{Note added.} While this manuscript was being completed, refs.~\cite{MSY,BajajLin} appeared, which also discuss the ground state and other aspects of the $SU(2)$ BFSS model.

\SZtocoff
\begin{acknowledgments}
We thank Sam Christian, Francisco Divi, Davide Gaiotto, Boris Khanikati, Rob Myers, Suriyah Rajalingam Kannagi, Jaroslav Scheinpflug and Andy Strominger for helpful discussions. We are especially grateful to Kabir Bajaj and Henry Lin for sharing with us their unpublished matrix bootstrap bounds. We used Claude Code (Opus and Fable) extensively at every stage of the project. We take full responsibility for the accuracy and integrity of the manuscript.
Research at Perimeter Institute is supported in part by the Government of Canada through the Department of Innovation, Science and Economic Development and by the Province of Ontario through the Ministry of Colleges and Universities.
\end{acknowledgments}
\twocolumngrid

\SZbibfont
\bibliography{References}

\SZtocon
\onecolumngrid
\clearpage
\section*{Supplemental Material}
\appendix
\renewcommand{\theHequation}{\thesection.\arabic{equation}}
\input{Sections/S.Supplemental}

\end{document}

%% file: Sections/0.Abstract.tex
We numerically solve for the ground state of the $SU(2)$ BFSS matrix model for M-theory. This state corresponds to the zero-energy bound state of two D0-branes or, by the BFSS conjecture, to the eleven-dimensional graviton with two units of momentum along the M-theory circle. The wavefunction exhibits a strongly coupled core, as well as an asymptotic power-law tail where the D0-branes are well separated. The coefficient of this power-law decay is the three-graviton coupling, previously computed by Herderschee and Maldacena, which our numerical solution reproduces with high accuracy. We also show that this coupling is protected by supersymmetry in the matrix model, and derive its value from the non-renormalized one-loop interaction of the two D0-branes. Lastly, we provide a simple representation-theoretic argument regarding the recently proposed excited bound states of the model.

%% file: Sections/1.Introduction.tex
\section{Introduction}\label{sec:intro}

The low-energy dynamics of open strings stretching between $N$ D0-branes in Type IIA string theory is famously given by the supersymmetric quantum mechanics with Hamiltonian~\cite{Claudson:1984th,deWit:1988wri}
\begin{equation}
H=\Tr\Big(\tfrac12 \big(P^{I}\big)^2-\tfrac{g^2}4[X^I,X^J]^2-\tfrac g2\psi^T\gamma^I[X^I,\psi]\Big) \ ,
\label{eq:H}
\end{equation}
where $X^I,\, I=1,\dots, 9$ are $N\times N$ Hermitian matrices describing the directions perpendicular to the branes, with conjugate momenta $P^I$, $\psi_\alpha,\,\alpha=1,\dots,16$ their Majorana fermionic superpartners and $\gamma^I$ the gamma matrices of $Spin(9)$. The Hamiltonian~\eqref{eq:H} can be obtained by dimensionally reducing ten-dimensional $\mathcal N=1$ super Yang--Mills theory to $0+1$ dimensions. According to the Banks--Fischler--Shenker--Susskind (BFSS) conjecture~\cite{BFSS}, in a particular limit, this matrix quantum mechanics is supposed to capture the flat-space $S$-matrix of eleven-dimensional M-theory. More precisely, the physical states of the model are $SU(N)$ singlets, corresponding to M-theory compactified on a lightlike circle with $N$ units of momentum along that circle~\cite{Susskind,Seiberg,Sen}. The eleven-dimensional graviton with $N$ units of momentum must then be a normalizable state of the $SU(N)$ BFSS matrix quantum mechanics~\cite{WittenDynamics,Witten95}, and graviton scattering corresponds to the scattering of these bound states.

The simplest such bound state to calculate in the matrix model is the zero-energy ground state of $SU(2)$ BFSS, the threshold bound state of two D0-branes. For the $SU(2)$ model, it is known through subtle Witten index computations~\cite{Yi,SethiStern} that the ground state exists and, together with the $Spin(9)$ invariance of every zero-energy state, that it is unique~\cite{SethiStern2,HaslerHoppe02}.\footnotemark[1]{\let\thefootnote\relax\footnotetext{\rule{0pt}{2.2em}The code and the data that reproduce our numerical results are available at \url{https://github.com/ssanjurjo/su2-bfss}.}} Due to the complexity of the Hamiltonian~\eqref{eq:H} and the fact that the system is strongly coupled at low energies, however, explicitly determining the ground-state wavefunction is a daunting task (see refs.~\cite{Wosiek05,Lundholm,HynekTrzetrzelewski,MichishitaTrzetrzelewski,LinYin} for earlier approaches). For instance, expanding the state in the Fock space basis of the $24$ complex fermions, the wavefunction is \textit{a priori} the solution of a system of $2^{24}$ coupled PDEs in $27$ bosonic variables!

In this article, we present a numerical solution for the ground state of the $SU(2)$ BFSS matrix model and determine its tail at large separation.\footnotetext[1]{For higher values of $N$, the index has been argued to equal one in refs.~\cite{PorratiRozenberg,GreenGutperle,KNS,MNS,KacSmilga}, and a proof of existence and uniqueness for every $N$ has been recently proposed in ref.~\cite{OpenAIunique}.}\setcounter{footnote}{1} The main simplification comes from imposing that the ground state should be a $Spin(9)\times SU(2)$ singlet. The underlying group theory then reduces this highly non-trivial PDE problem to a system of $2066$ coupled PDEs in just three bosonic variables, which is numerically tractable. We emphasize that this enormous reduction is special to $SU(2)$ and is not expected for higher $N$.

Using the variational method, we are able to obtain the ground-state wavefunction and estimate that its overlap with the true zero-energy state of eq.~\eqref{eq:H} is at least $99.8\%$. From our numerical solution, we can infer various properties about the wavefunction as a function of $R$, where $R$ corresponds to the relative separation of the two D0-branes along the flat directions of the potential in eq.~\eqref{eq:H}. The wavefunction exhibits two main features: a strongly coupled core region for small $R$ where the matrices do not commute and the majority of the probability is concentrated, and an asymptotic tail for large $R$ that describes the relative motion of the D0-branes. The tail enjoys a power-law decay of the form $CR^{-9}$, which is dictated by supersymmetry and $Spin(9)$ invariance~\cite{HalpernSchwartz,GrafHoppe,FGHHY}. We show that the coefficient $C$ of this decay is the three-graviton coupling in eleven dimensions. On the supergravity side $C$ is fixed by the eleven-dimensional super-Poincaré symmetry, but requires a non-trivial computation in the matrix model that was undertaken in ref.~\cite{HM3}. Our numerical solution reproduces this value to within $0.1\%$. Moreover, to probe aspects of the wavefunction's core, we evaluate moments, \textit{e.g.} $\langle\Tr X^2\rangle$, which we find to be in excellent agreement with the bounds that the matrix bootstrap~\cite{HHK,Lin23,LinZheng} gives for $SU(2)$.

As already mentioned, in supergravity, the three-graviton coupling is fixed by supersymmetry and eleven-dimensional Lorentz invariance. The matrix model, on the other hand, does not realize Lorentz invariance manifestly. Nonetheless, by relating the amplitude to a supersymmetric index, Herderschee and Maldacena found that the matrix model reproduces the supergravity coupling at every finite $N$~\cite{HM3}, although the BFSS conjecture only requires agreement at large $N$. We explain why this is the case in the $SU(2)$ model. In particular, two gravitons scatter in supermultiplets of the supercharges and $Spin(9)$, and the $S$-matrix acts on each multiplet as a single phase due to a super-Ward identity. The channel of the bound state shares its multiplet with a channel that contains no bound state, whose phase at low energy is fixed by the non-renormalized one-loop interaction of two D0-branes. Equating the two phases requires the bound state to cancel the spin-dependent part of this interaction, which fixes the tail coefficient, and hence the coupling, to its supergravity value.

On top of the ground state, it is known that the system enjoys a continuum of scattering states~\cite{dWLN}, and no other bound states were anticipated~\cite{BFSS,TASI}. Recently, however, a proof of the existence of positive-energy bound states of the $SU(2)$ model was proposed~\cite{OpenAIpos}. Here we provide a simple argument, purely based on supersymmetry and $Spin(9)$ representation theory, which explains the mechanism that prevents these normalizable states from decaying into the continuum. The main idea is that the scattering states of two D0-branes do not carry all possible $Spin(9)$ representations. Hence, the model can have bound states in the remaining representations, which do not decay, since no scattering state has the same quantum numbers. Supersymmetry, in addition, enables the existence of bound states also for representations that are carried by scattering states, whenever such a representation belongs to a multiplet that also contains a state that cannot be carried by gravitons. Curiously, we argue that this protection mechanism fails for $N\geq 5$, potentially suggesting the absence of excited bound states in these cases.

The rest of this article is organized as follows. In Section~\ref{sec:tail} we discuss various aspects of the ground-state wavefunction of $SU(2)$ BFSS, explain how its tail is related to the three-point amplitude and present our numerical solution. In Section~\ref{sec:ward} we derive the three-graviton coupling within the model by combining a super-Ward identity with the one-loop interaction of the D0-branes. In Section~\ref{sec:excited} we describe some representation-theoretic aspects of bound states above the zero-energy ground state, and we conclude in Section~\ref{sec:outlook} with open directions. Various details are contained in the Appendices. Appendix~\ref{app:conventions} contains our conventions, Appendix~\ref{ap:three_grav_coupling} the conversion of the three-point coupling computed in ref.~\cite{HM3} to our conventions, Appendix~\ref{app:reduction} the singlet sector reduction of the ground-state problem, Appendix~\ref{app:numerics} further details on the derivation and the properties of the numerical solution, and Appendix~\ref{app:phase} the computation of the phase shifts relevant for Section~\ref{sec:ward}.

%% file: Sections/2.BoundState.tex
\section{The ground state}\label{sec:tail}

In this section we discuss properties of the ground state of $SU(2)$ BFSS and verify them against a direct numerical solution for the wavefunction.

\subsection{General features of the model}

We work in units where $g=2$ and, following ref.~\cite{FGHHY}, write the Hamiltonian~\eqref{eq:H} in the $SU(2)$ basis as
\begin{equation}\label{eq:SU2_H}
H=\sum_I\vec p_I\cdot\vec p_I+\sum_{I<J}|\vec q_I\times\vec q_J|^2+i\sum_{I,\alpha,\beta}\gamma^I_{\alpha\beta}\vec q_I\cdot(\vec\Theta_\alpha\times\vec\Theta_\beta) \ ,
\end{equation}
where $X^I=q_{IA}t^A$, $P^I=2p_{IA}t^A$ and $\psi_\alpha=\sqrt2\Theta_{\alpha A}t^A$, with $t^A=\sigma^A/2$ (see Appendix~\ref{app:conventions}). The bosonic potential vanishes on the valley $\vec q_I=r_I\vec n$, where the matrices commute. Here $r\in\R^9$ is the relative position of the two D0-branes, $R=|r|$ their separation, and $\vec n$ a unit vector in color space. We will repeatedly use three properties of the valley.

\textit{Flatness.} At large $R$, the components transverse to $\vec n$, which describe the strings stretched between the two D0-branes, are bosonic and fermionic oscillators of frequency $2R$ whose zero-point energies cancel exactly. The relative motion is, therefore, free and governed by the Hamiltonian $H_{\rm rel}=k^2$, where $k_I=-i\partial/\partial r_I$ denotes the momentum conjugate to $r$. As a result, the spectrum is continuous~\cite{dWLN} and, since $H\geq0$, it extends over $[0,\infty)$.

\textit{Polarizations.} The sixteen fermions along the valley, $\theta_\alpha=\vec n\cdot\vec\Theta_\alpha$, remain massless and satisfy $\{\theta_\alpha,\theta_\beta\}=\delta_{\alpha\beta}$. They act on $2^8$ states, which form the representation $\rep{256}=\rep{44}\oplus\rep{84}\oplus\rep{128}$ of $Spin(9)$. We label representations by their highest weights, so that $\rep{44}=(2,0,0,0)$, $\rep{84}=(1,1,1,0)$ and $\rep{128}=(\tfrac32,\tfrac12,\tfrac12,\tfrac12)$. Together with the $\rep{256}$ of the decoupled center of mass, these are the polarizations of two graviton multiplets in eleven-dimensional supergravity, so that at large separation the relative motion of two D0-branes is that of two free gravitons.

\textit{Exchange rule.} Exchanging the two D0-branes is the Weyl reflection of $SU(2)$, which reverses $\vec n$. It acts as $r\to-r$ and $\theta\to-\theta$, and leaves the vacuum of the stretched strings invariant. On the $\rep{256}$, $\theta\to-\theta$ is the fermion parity, which equals $+1$ on $\rep{44}\oplus\rep{84}$ and $-1$ on $\rep{128}$. Since the Weyl reflection is a gauge rotation by $\pi$, the Gauss constraint requires physical states to be invariant under it. The partial wave $(\ell,0,0,0)$ can therefore only carry
\begin{equation}\label{eq:exchange}
\begin{aligned}
&(\ell,0,0,0)\otimes(\rep{44}\oplus\rep{84})\quad\text{for even }\ell \ ,\\
& (\ell,0,0,0)\otimes\rep{128}\qquad\quad \ \ \text{for odd }\ell \ .
\end{aligned}
\end{equation}

In addition to this continuum of scattering states, the $SU(2)$ model has a normalizable zero-energy bound state at threshold. Its existence was established by the computation of the Witten index, which was shown to equal one~\cite{Yi,SethiStern}. Since the index counts bosonic minus fermionic zero-energy states, it does not by itself imply uniqueness. However, every zero-energy state is invariant under $Spin(9)$~\cite{SethiStern2,HaslerHoppe02}. In particular, it is invariant under $-\mathbf 1\in Spin(9)$, which acts as $(-1)^F$, so it is bosonic, and the index then implies that the ground state is unique~\cite{SethiStern2}. Its asymptotic form was determined in refs.~\cite{HalpernSchwartz,GrafHoppe,FGHHY}, as we review below. For general $N$, a proof of existence and uniqueness of the ground state was recently proposed in ref.~\cite{OpenAIunique}. As we will describe, together with gauge invariance, these properties lead to a substantial simplification of the problem of determining the wavefunction.

\subsection{The tail of the ground-state wavefunction and the three-graviton coupling}\label{sec:vertex}

At large separation, the stretched strings constitute fast degrees of freedom, whereas the relative position $r$ and the zero modes $\theta_\alpha$ are slow, and an asymptotic analysis determines the form of the ground state~\cite{HalpernSchwartz,GrafHoppe,FGHHY}. Let $y\in\R^{16}$ denote the physical components of $\vec q_I$ transverse to $\vec n$. As explained above, the strings are free oscillators, and thus occupy the normalized ground state $\chi_R(y)=(R/\pi)^4e^{-R|y|^2/2}$, of width $R^{-1/2}$, together with the corresponding fermionic vacuum $|\Omega_R\rangle$. As a consequence of the flatness of the valley, to leading order, the slow sector is governed by the free supercharges,
\begin{equation}\label{eq:freeQ}
Q_\alpha\simeq\sqrt2\gamma^I_{\alpha\beta}k_I\theta_\beta \ , \qquad \{Q_\alpha,Q_\beta\}\simeq2\delta_{\alpha\beta}k^2 \ ,
\end{equation}
so that a zero-energy state behaves, asymptotically, as
\begin{equation}
|\Psi\rangle\simeq \chi_R(y)|\Omega_R\rangle\otimes\Psi_{\rm rel}(r) \ ,
\end{equation}
where $\Psi_{\rm rel}$ takes values in the $\rep{256}$ and has harmonic components that decay at infinity (as implied by annihilation by the supercharges and required by normalizability, respectively).

The symmetries of the model determine $\Psi_{\rm rel}$ uniquely. First, a decaying harmonic function in the partial wave $(\ell,0,0,0)$ scales as $R^{-(\ell+7)}$. Now, $Spin(9)$ invariance requires the orbital and polarization representations to coincide. Among the combinations permitted by the exchange rule~\eqref{eq:exchange}, the only combination that fills the bill is $(2,0,0,0)\otimes\rep{44}$~\cite{Polchinski99}. Put differently, we have\footnote{Taking two derivatives of the nine-dimensional Green's function $R^{-7}$ produces the required $\ell=2$ harmonic.}
\begin{equation}\label{eq:tail}
\Psi_{\rm rel}(r)\propto\partial_I\partial_J\frac{1}{R^7}|h_{IJ}\rangle\simeq \frac{C}{R^9}\Big(\hat r_I\hat r_J-\frac19\delta_{IJ}\Big)|h_{IJ}\rangle \ ,
\end{equation}
where $C$ is a constant that the asymptotic analysis leaves undetermined, $\hat r=r/R$ and $|h_{IJ}\rangle$ denote the symmetric traceless states of the $\rep{44}$, normalized so that $\langle h_{IJ}|h_{KL}\rangle=\tfrac12(\delta_{IK}\delta_{JL}+\delta_{IL}\delta_{JK})-\tfrac19\delta_{IJ}\delta_{KL}$. One can verify that this state is indeed annihilated by the free supercharges~\cite{GrafHoppe,FGHHY,HaslerHoppe}.

As expected for a state at threshold, the wavefunction exhibits power-law, rather than exponential, decay. In addition, the probability density for the separation of the two D0-branes behaves as
\begin{equation}\label{eq:PR}
P(R)=R^8\int d\Omega_8|\Psi_{\rm rel}|^2\simeq \frac 89 \Omega_8 |C|^2R^{-10} \ ,
\end{equation}
where $d\Omega_8$ is the area element of the unit eight-sphere. The coefficient $C$ depends on the full wavefunction. In fact, it is fixed by the three-graviton coupling of eleven-dimensional supergravity, as we now turn to discuss.

It is straightforward to show that $C$ corresponds to a physical amplitude. In short, a power-law tail in position space corresponds to a pole at $k^2=0$ in momentum space, and the residue of this pole is the amplitude for the bound state to split into its constituents. To be more precise, we denote the ground state by $\ket{B}$, write $\Psi_{\rm rel}(r)=\int\frac{d^9k}{(2\pi)^9}e^{ik\cdot r}\widetilde\Psi_{\rm rel}(k)$, and use that, away from the origin,
\begin{equation}
\begin{aligned}
G_9(r)&=\int\frac{d^9k}{(2\pi)^9}\frac{e^{ik\cdot r}}{k^2}=\frac{15}{32\pi^4R^7} \ ,\\
\partial_I\partial_JG_9(r)&=\frac{9}{\Omega_8}R^{-9}\Big(\hat r_I\hat r_J-\frac{\delta_{IJ}}{9}\Big) \ ,
\end{aligned}
\label{eq:G9}
\end{equation}
where $G_9$ is the nine-dimensional Green's function and $\Omega_8=32\pi^4/105$. Comparing with eq.~\eqref{eq:tail}, the tail corresponds in momentum space to
\begin{equation}
\widetilde\Psi_{\rm rel}(k)\simeq-C\frac{\Omega_8}{9}\frac{k_Ik_J}{k^2}\ket{h_{IJ}} \ ,
\end{equation}
up to contributions that are less singular at $k=0$ and correspond to faster-decaying terms in position space. On the other hand, the relative Hamiltonian along the valley is $k^2$ plus an interaction $V$, so that the zero-energy state $\ket{B}$ satisfies $\bra{k,a}V\ket{B}=-k^2\langle k,a|B\rangle$. Here, $|k,a\rangle=|k\rangle \otimes |a\rangle$ are the eigenstates of the free relative Hamiltonian normalized as $\langle k',a'|k,a\rangle=(2\pi)^9\delta^9(k'-k)\delta_{a'a}$, where $a$ labels the relative polarization. Therefore, the residue of the pole yields the amplitude
\begin{equation}
\bra{k,a}V\ket{B}=C\frac{\Omega_8}{9}k_Ik_J\langle a|h_{IJ}\rangle \ ,
\label{eq:split-vertex}
\end{equation}
which is quadratic in the relative transverse momentum $k$ and to which only the component of the polarization $a$ in the $\rep{44}$ contributes. This is the structure of the cubic vertex of eleven-dimensional supergravity, so $C$ plays the role of the three-graviton coupling, which we determine below by comparison with supergravity~\cite{HM3}. On shell, eq.~\eqref{eq:split-vertex} is the amplitude for a graviton with two units of light-cone momentum to split into two gravitons with one unit each. For real momenta the on-shell condition $k^2=0$ implies $k=0$, where the amplitude vanishes, consistent with $\ket{B}$ being a bound state. The amplitude is nonvanishing for complex momenta, where it is defined by analytic continuation.

On the gravity side, the three-point amplitude is fixed by supersymmetry and eleven-dimensional Lorentz invariance, up to an overall factor proportional to the gravitational coupling $\kappa_{11}$, with $\kappa_{11}^2=16\pi G_{11}$. Since the matrix model does not manifestly realize all of these symmetries, reproducing this amplitude requires a non-trivial computation. In ref.~\cite{HM3}, Herderschee and Maldacena related this computation to a supersymmetric index and showed that the matrix model reproduces the supergravity amplitude even at finite $N$. Expressing the explicit value of the amplitude from ref.~\cite{HM3} in our conventions (see Appendix~\ref{ap:three_grav_coupling}), we arrive at
\begin{equation}\label{eq:tail_three_grav_coupling_prediction}
|C|=\frac{945\sqrt2}{8\pi^2} \ , \qquad P(R)\simeq\frac{7560}{R^{10}} \ ,
\end{equation}
where we also substituted the value of $C$ in the probability density~\eqref{eq:PR}. Lastly, let us note that the subleading terms of the tail of the wavefunction probe the interaction of the two gravitons at large separation. Determining their powers and coefficients requires a further asymptotic calculation, which we do not attempt here.

\subsection{The numerical solution}\label{sec:solved}

\paragraph{Setup.} We now turn to the numerical determination of the ground state of the $SU(2)$ model. The ground state is the normalizable solution of $H\ket{\Psi}=0$, with $H$ the Hamiltonian~\eqref{eq:H}. For $SU(2)$, there are $27$ bosonic coordinates $q_{IA}$ and $48$ real fermions $\Theta_{\alpha A}$, which combine into $24$ complex fermions with a Fock space of dimension $2^{24}$. In a basis $\ket{s}$ of this space, the state reads
\begin{equation}
\ket{\Psi}=\sum_{s=1}^{2^{24}}\Psi_s(q)\ket{s} \ .
\end{equation}
As discussed in the previous subsections, the ground state is moreover a $Spin(9)\times SU(2)$ singlet. As detailed in Appendix~\ref{app:reduction}, these constraints determine the wavefunction everywhere from its values on the configurations
\begin{equation}
q=\begin{pmatrix}\mathrm{diag}(r_1,r_2,r_3)\\0_{6\times3}\end{pmatrix} \ , \qquad r_1\geq r_2\geq r_3\geq0 \ ,
\end{equation}
to which any $q$ can be brought by a $Spin(9)\times SU(2)$ transformation, so that the $r_i$ are the singular values of the $9\times3$ matrix $q$. On these configurations the wavefunction takes values in the $2066$-dimensional subspace of the Fock space that is invariant under their stabilizer $SU(4)\times\mathbb Z_2\times\mathbb Z_2$. Acting on such wavefunctions, the Hamiltonian reduces to the operator derived in eq.~\eqref{eq:Hred},
\begin{equation}\label{eq:Hred_main-text}
\begin{aligned}
H_{\rm red}=&-\sum_i\frac1J\frac{\partial}{\partial r_i}\Big(J\frac{\partial}{\partial r_i}\Big)\\
&+\sum_{i<j}\Bigg[\frac{K_{S,ij}^2}{2(r_i-r_j)^2}+\frac{K_{D,ij}^2}{2(r_i+r_j)^2}\Bigg]\\
&+\sum_i\frac{O_i}{r_i^2}+\sum_{i<j}r_i^2r_j^2+\sum_ir_iY_i \ ,
\end{aligned}
\end{equation}
where $J=\prod_{i<j}|r_i^2-r_j^2|\prod_ir_i^6$ is the volume factor of the reduction, and the fermionic operators $K_{S,ij}$, $K_{D,ij}$, $O_i$ and $Y_i$, which act within the $2066$-dimensional space, are given in eqs.~\eqref{eq:red1}, \eqref{eq:red2} and~\eqref{eq:red3}. The ground state is therefore the solution of $H_{\rm red}\ket{\Psi}=0$, normalizable with respect to the measure $Jd^3r$, of the form
\begin{equation}
\ket{\Psi}=\sum_{\tilde s=1}^{2066}\Psi_{\tilde s}(r_1,r_2,r_3)\ket{\tilde s} \ .
\end{equation}
This is a system of $2066$ coupled second-order partial differential equations in $r_1$, $r_2$ and $r_3$. From the explicit reduction presented in Appendix~\ref{app:reduction}, it follows that this impressive reduction in the number of independent variables is a distinctive feature of the $SU(2)$ model, which does not generalize to higher values of $N$.

\begin{figure*}[tbp]
\centering
\includegraphics[width=\textwidth]{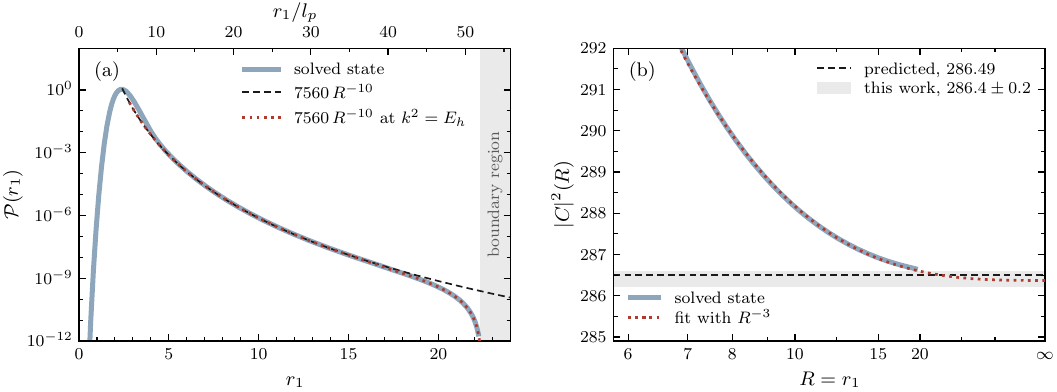}
\caption{Plot of the numerical solution for the ground state. (a) Probability density of the largest singular value $r_1$, with the tail $7560\,R^{-10}$ at zero energy and at the energy of the state in the finite box. (b) The value of the tail coefficient $|C|^2$ that the density gives at each $R$, with its far-field fit. The band corresponds to the result~\eqref{eq:Cdirect} and the horizontal dashed line to the prediction~\eqref{eq:tail_three_grav_coupling_prediction}.}
\label{fig:density}
\end{figure*}

\paragraph{Method.} We solve the reduced problem by the Galerkin variational method, in the domain $r_1\geq r_2\geq r_3\geq0$. After an integration by parts, which produces no boundary terms since $J$ vanishes on the walls of the domain, the energy of a state takes the form
\begin{equation}
\langle\Psi|H_{\rm red}|\Psi\rangle=\int d^3r J\Big[\sum_i\big\|\partial_{r_i}\Psi\big\|^2+\Psi^\dagger A(r)\Psi\Big] \ ,
\label{eq:form}
\end{equation}
where $A(r)$ denotes the terms of eq.~\eqref{eq:Hred_main-text} without derivatives. We then minimize the ratio $\langle\Psi|H_{\rm red}|\Psi\rangle/\langle\Psi|\Psi\rangle$ over a finite space of trial functions, which amounts to a generalized eigenvalue problem for the coefficients of the trial state. The trial functions are determined by the following three features of the problem: the coefficients of eq.~\eqref{eq:Hred_main-text} are singular on the walls of the domain, the flat valley must be followed far while the state becomes ever narrower across it, and the state has $2066$ coupled fermionic components. We will now offer an overview of our numerical strategy, and defer further details to Appendix~\ref{app:numerics}.

We use the coordinates $a=r_3$, $b=r_2-r_3$ and $c=r_1-r_2$, in which the domain becomes the octant $a,b,c\geq0$ and its walls become coordinate planes. We restrict to the box $a,b\leq3$, $c\leq L_c$, with $L_c=15$ or $22$, and impose Dirichlet conditions on its outer faces. We then cover the box with a rectangular grid of cells. On each cell the spatial basis functions are polynomials in $a$, $b$ and $c$, fixed by their values at a lattice of equally spaced points of the cell, the nodes, which include its corners. Neighboring cells share the nodes on their common face, so that the functions are continuous. We then expand the state as
\begin{equation}\label{eq:ansatz}
\Psi_h(r)=\eta(r)\sum_nN_n(r)\sum_mx_{nm}u_{n,m} \ ,
\end{equation}
where each basis function $N_n$ equals one at the node $r_n$ and vanishes at all other nodes, and $u_{n,m}$ are a few vectors in the $2066$-dimensional fermionic space attached to each node $n$, the local channels labeled by $m$. The envelope $\eta=\exp[-(r_1r_2^2+r_1r_3^2+r_2r_3^2)/2]$ is the ground state of the transverse oscillators along the valley, chosen so that the remaining wavefunction $\Psi_h/\eta$ varies slowly across the valley. Lastly, the complex coefficients $x_{nm}$ are the variational parameters.

In the spirit of the Born--Oppenheimer approximation, the initial fermionic channels are taken to be the eight lowest eigenvectors of $A(r)$ averaged over the node with weight $J\eta^2N_n^2$. In particular, along the valley the lowest of them is the fermionic ground state, whose energy $-16R$ cancels the zero-point energy $16R$ of the transverse oscillators. Subsequently, the channels are enriched iteratively. After solving the variational problem with the current fermionic channels, we evaluate the residual $(H_{\rm red}-E_h)\Psi_h$ on the full $2066$-dimensional space, and at every node we essentially add the component orthogonal to the existing channels (for more details see Appendix~\ref{app:numerics}). In practice, ten rounds give up to $28$ channels per node, after which the energy saturates.

\paragraph{Results.} Figure~\ref{fig:density} summarizes the numerical solution. Our best state uses quartic polynomials in $a$ and $b$ on the $10\times10$ mesh across the valley, with $L_c=22$, and has $E_h=0.0003$. Since $\Psi_h$ is an approximation, it is not annihilated by the supercharges, and $E_h=\|Q_\alpha\Psi_h\|^2$ for every $\alpha$ measures the residual violation of supersymmetry. Up to quadrature errors, $E_h$ is an upper bound on the lowest energy in the box, and it decreases as the trial space is enlarged. Expanding $\Psi_h$ in the eigenstates of the box, whose two lowest levels are $E_0$ and $E_1$, its squared overlap with the true ground state $\Psi_0$ of the box obeys
\begin{equation}\label{eq:overlap}
|\langle\Psi_0|\Psi_h\rangle|^2\geq\frac{E_1-E_h}{E_1-E_0}\geq1-\frac{E_h}{E_1} \ .
\end{equation}
With the computed $E_1=0.1742$, we find the right-hand side to be about $0.998$. The state also satisfies the virial relations $\vev{T}=\vev{V_B}=-\tfrac12\vev{V_F}$ of the exact zero-energy state, which follow from the scaling of the kinetic, quartic and Yukawa terms under $q\to\lambda q$~\cite{Polchinski99}. For our best state we find $\vev{T}=27.166$, $\vev{V_B}=27.164$ and $\vev{V_F}=-54.329$, so that $\vev{T}$ and $\vev{V_B}$ agree to $0.01\%$. However, since the continuum starts at zero energy, a small $E_h$ controls neither the far field nor the moments that weigh it, and their errors must be determined by refinement (as further explained in Appendix~\ref{app:numerics}).

\textit{Size of the state.} Figure~\ref{fig:density}(a) shows the marginal density~\eqref{eq:full_PR} of the largest singular value, from the core to the (numerical) wall of the box. Along the valley, it is the density~\eqref{eq:PR} of the separation of the two D0-branes. In the strongly coupled core, where the matrices do not commute, this quantity remains well defined. It peaks at $r_1\simeq5.5\lp$. The root-mean-square hyperradius is $\vev{\rho^2}^{1/2}=7.4\lp$, where $\rho^2=\sum_{I,A}q_{IA}^2=2\Tr X^IX^I$ is normalized so that $\rho$ coincides with $R$ along the valley. Only $0.4\%$ of the probability lies beyond $R\simeq4$, \textit{i.e.} $9.3\lp$, which is where the valley begins.

\textit{Tail coefficient.} Along the valley, $\sqrt{\mathcal P}$ obeys the free radial equation~\eqref{eq:radial_app} of an $\ell=2$ wave in nine dimensions. Since $\Psi_h$ is a variational state in a finite box, its energy $E_h$ is small but nonzero. Its decaying solution then deviates from the tail~\eqref{eq:tail} by a relative amount $E_hR^2/9$ in the density. We therefore fit $\sqrt{\mathcal P}$ to the exact solution at energy $E_h$, including the reflection off the wall of the box and corrections of relative order $R^{-3}$, and continue its decaying branch analytically to zero energy, which gives $|C|^2$ (Appendix~\ref{app:numerics}). A scan over $k^2$ returns $k^2=E_h$ to within a few percent, \textit{i.e.} the far field is the free $d$-wave at the energy of the state, further justifying our numerics. Since $|C|^2$ is not variational, and different refinements change it at different rates per unit of $E_h$, we do not extrapolate it in $E_h$. Instead, we fit the far field of our best state, and estimate the uncertainty from the shifts measured when each of the remaining discretization parameters is refined separately. We find
\begin{equation}
|C|^2=286.4\pm0.2 \ ,
\label{eq:Cdirect}
\end{equation}
where the uncertainty is a systematic estimate, given by the largest of the refinement shifts. Comparing this value to the three-graviton coupling of M-theory~\eqref{eq:tail_three_grav_coupling_prediction}, $|C|^2\simeq286.49$, we observe that the two values differ by less than $0.1\%$, within the uncertainty. To investigate the agreement further, Figure~\ref{fig:density}(b) shows the local coefficient, \textit{i.e.} the value of $|C|^2$ that the density gives at each $R$.

\textit{Moments.} The core of the state is characterized by the moments of the invariants
\begin{equation}
\begin{gathered}
t_n=\Tr\big[(q^Tq)^n\big]=\sum_ir_i^{2n} \ ,\\
\Tr X^{2n}\equiv\Tr(X^IX^I)^n=2^{1-2n}\,t_1^n \ ,
\end{gathered}
\end{equation}
where $q$ is the $9\times3$ matrix of coordinates. The first of them is the squared hyperradius, $t_1=\sum_{I,A}q_{IA}^2=2\Tr X^IX^I$. Since $X^IX^I=\tfrac14t_1\mathds 1$ for $SU(2)$, the moments $\vev{\Tr(X^IX^I)^n}$ used in the matrix bootstrap~\cite{HHK,Lin23,LinZheng} are proportional to $\vev{t_1^n}$. Moreover, the bosonic potential is $V_B=\tfrac12(t_1^2-t_2)$, and in the exact state its expectation value equals the kinetic energy $\vev{T}$, by the virial relations. Along the valley, $t_1^n$ grows as $R^{2n}$, so that the tail~\eqref{eq:tail} makes $\vev{t_1^n}$ finite only for $n\leq4$. In particular, $\vev{\Tr(X^IX^I)^n}$ diverges for $n\geq5$~\cite{Polchinski99}. The moments of our numerical solution are collected in Table~\ref{tab:moments}. The obtained moments appear to lie within the tight bounds of the $SU(2)$ matrix bootstrap.\footnote{We thank Sam Christian for drawing our attention to these bounds. We are grateful to Kabir Bajaj and Henry Lin for sharing with us their unpublished matrix bootstrap results for $SU(2)$ BFSS. Note that the matrix bootstrap literature typically uses different conventions than the ones we have adopted here. The conversion between the two is explained in Appendix~\ref{app:numerics}, with Table~\ref{tab:obs} providing our moments in both normalizations.}

\begin{table}[ht]
\centering
\begin{tabular}{@{}lr@{}}
\toprule
moment & value\\
\midrule
$\vev{\Tr X^2}$ & $5.0865(4)$\\
$\vev{\Tr X^4}$ & $13.643(3)$\\
$\vev{\Tr X^6}$ & $38.90(1)$\\
$\vev{\Tr X^8}$ & $126.09(8)$\\
$\vev{V_B}$ & $27.164(3)$\\
\bottomrule
\end{tabular}
\caption{Moments of the ground state for $g=2$ in eq.~\eqref{eq:H}, with $\Tr X^{2n}\equiv\Tr(X^IX^I)^n$. The last row equals the kinetic energy $\vev{T}$ by the virial relations.}
\label{tab:moments}
\end{table}

%% file: Sections/3.Ward.tex
\section{The three-graviton coupling from supersymmetry}\label{sec:ward}

As already remarked, the evaluation of the three-point amplitude in the matrix model is non-trivial. As noticed in ref.~\cite{HM3}, the value of this amplitude coincides with the three-graviton coupling of eleven-dimensional supergravity even at finite values of $N$. That is slightly surprising, since the BFSS conjecture demands agreement between the two only at infinite $N$. In this section we offer an alternative derivation of the coupling, within $SU(2)$ BFSS. We show, in particular, that the value of the amplitude is fixed to be the three-graviton coupling via a super-Ward identity. We now outline the argument, while presenting the explicit details in Appendix~\ref{app:phase}.

The argument rests on the fact that supersymmetry assigns the same phase shift to all channels of two-graviton scattering that belong to the same supermultiplet. At low energies, the phase shift of a $d$-wave channel is determined, at leading order in the momentum, by the long-range interaction of two separated D0-branes (see eq.~\eqref{eq:bornkernel}), which is known~\cite{DKPS,PSW98,NicolaiPlefka} and is not renormalized~\cite{PSS}. Its spin-dependent terms contribute differently to different channels. In the singlet channel, which contains the ground state, the bound state contributes at leading order with a weight proportional to $|C|^2$. Requiring, as the super-Ward identity demands, that the phase shift of the singlet coincide with that of a channel in the same supermultiplet that contains no bound state then fixes $|C|^2$ to the value of eq.~\eqref{eq:tail_three_grav_coupling_prediction}, \textit{i.e.} to the three-graviton coupling of eleven-dimensional supergravity.

To find such a channel, note that all states with $E>0$ organize into supermultiplets with representations $W\otimes\mathbf{256}$, where the supercharges act trivially on $W$. Now, positive-energy states with two gravitons further have $W=(2m,0,0,0),\, m\geq0$. To see why, recall that on such states the supercharges reduce to the free ones of eq.~\eqref{eq:freeQ}. Fixing the direction of relative momentum $\hat k$, the matrix $\gamma\!\cdot\!\hat k$ that appears in them is a rotation of the sixteen $\theta_\beta$, which can be undone by a unitary operator acting on the polarizations. After this redefinition the supercharges no longer depend on the direction $\hat k$, so that $W$ consists of functions of $\hat k$. Moreover, the unitary operators at $\hat k$ and $-\hat k$ differ by the fermion parity, and hence the exchange of the two D0-branes acts on these functions only through $\hat k\to-\hat k$, selecting the harmonics of a definite parity. The parity is even, since the states of eq.~\eqref{eq:exchange} contain a singlet, which only the harmonic $\ell=2$ can form with the $\rep{256}$. We conclude that, for the two-graviton scattering states, $W=\oplus_{m\geq0}(2m,0,0,0)$, \textit{i.e.} it contains all even-rank symmetric traceless tensors of $SO(9)$ with multiplicity one.

Consider now the scattering of two gravitons in the singlet of $Spin(9)$ and in the sector $\rep{495}=(2,2,0,0)$.\footnote{Among the sectors of the multiplet $(2,0,0,0)\otimes\rep{256}$, the $\rep1$, the $\rep{36}$ and the $\rep{495}$ contain a single two-graviton channel; the $\rep{36}$ would serve equally well.} In both sectors the two gravitons appear only once, as a $d$-wave with polarization $\rep{44}$, and both channels belong to the multiplet $(2,0,0,0)\otimes\rep{256}$, since $\rep{44}\otimes\rep{44}\supset\rep1\oplus\rep{495}$. Their phase shifts are related by the super-Ward identity, which expresses the invariance of the $S$-matrix under rotations and under the asymptotic supercharges, $SQ_\alpha(k_i)=Q_\alpha(k_f)S$. At fixed energy $E=k^2$, the sixteen operators $Q_\alpha/\sqrt E$ generate a Clifford algebra, whose only irreducible module is the $\rep{256}$. By Schur's lemma, an operator that commutes with them acts on $\bigoplus_mW_m\otimes\rep{256}$ as $X\otimes\mathds 1$, and if it also commutes with rotations, $X$ is a multiple of the identity on each $W_m=(2m,0,0,0)$, since these representations are inequivalent and appear once. Hence the $S$-matrix is a single phase on each multiplet, and in particular
\begin{equation}
\delta_{\rep1}(k)=\delta_{\rep{495}}(k) \ .
\label{eq:equalphases}
\end{equation}
This conclusion requires all sixteen supercharges, since under any proper subset of them the $\rep{256}$ is reducible, and the $S$-matrix could then act on its irreducible pieces with different phases. The singlet is the channel of the bound state, whose tail~\eqref{eq:tail} is precisely a $d$-wave with polarization $\rep{44}$. The $\rep{495}$, on the other hand, contains no state of zero energy, since every such state is a singlet~\cite{SethiStern2,HaslerHoppe02}.

We compare the two phase shifts at low energy, where both can be computed, and this suffices to determine $|C|^2$. Two regions contribute to the phase shift of a $d$-wave. At long distances the interaction of the two D0-branes is known (see eq.~\eqref{eq:bornkernel}), and its first Born term contributes to $\tan\delta$ at order $k^9$, while its iterations and the higher loops enter at higher orders. At short distances, where the matrices do not commute, we assume that the dynamics acts as an interaction of short range~\cite{KabatPouliot,DKPS}. The centrifugal barrier keeps a $d$-wave of low energy away from this region, which therefore contributes only at order $k^{11}$, in line with the usual suppression $\tan\delta_\ell\propto k^{2\ell+1}$ in three dimensions~\cite{LandauLifshitz} and $k^{2\ell+7}$ in nine.

The singlet is an exception, since it contains a state of zero energy. Two gravitons in this channel can merge into the bound state and split again, which adds to their interaction the term (see Appendix~\ref{app:phase})
\begin{equation}
K_B=\frac1{k^2}\bra{k_f}V\ket{B}\bra{B}V\ket{k_i} \ .
\label{eq:KB}
\end{equation}
Each vertex (eq.~\eqref{eq:split-vertex}) is quadratic in $k$, but the intermediate state has zero energy, and the denominator $k^2$ lowers the order of this contribution from $k^{11}$ to $k^9$. Its coefficient is negative and proportional to $|C|^2$.

The long-range interaction is not renormalized~\cite{PSS,HKS}. It consists of the $v^4/r^7$ term~\cite{DKPS} and of four spin-dependent terms~\cite{PSW98,NicolaiPlefka} (see eq.~\eqref{eq:bornkernel}). Supersymmetry fixes the four spin-dependent contributions in terms of the first. At order $k^9$ we find (see Appendix~\ref{app:phase})
\begin{equation}
\begin{aligned}
\tan\delta_{\rep1}&\simeq\frac{k^9}{8!}+\frac8{945}k^9-\frac{256\pi^4}{945^3}|C|^2k^9 \ ,\\
\tan\delta_{\rep{495}}&\simeq\frac{k^9}{8!} \ ,
\end{aligned}
\label{eq:twophases}
\end{equation}
where the three terms of the first line come from the $v^4/r^7$ term, from the spin-dependent terms, which cancel in the $\rep{495}$, and from the bound state. The equality~\eqref{eq:equalphases} thus requires the bound state to cancel the spin-dependent terms of the singlet, which gives $|C|^2=893025/(32\pi^4)$, the value of eq.~\eqref{eq:tail_three_grav_coupling_prediction}.

%% file: Sections/4.Excited.tex
\section{Excited bound states}\label{sec:excited}

As discussed in Section~\ref{sec:tail}, the flat directions produce a continuum of scattering states above the zero-energy ground state. Accordingly, it was believed that the ground state is the only bound state of the model~\cite{BFSS,TASI} (see however ref.~\cite{DFS96}). Following the proposed proof of the existence of bound states for the $SU(2)$ model with $E>0$~\cite{OpenAIpos}, we would like to give a simple argument that explains how supersymmetry and $Spin(9)$ symmetry can protect excited bound states of two D0-branes from decaying into the continuum of scattering states.

The main point is that scattering states occupy only a small fraction of the possible $Spin(9)$ representations. Recall, in particular, that the scattering states come in the representations of eq.~\eqref{eq:exchange}. On the other hand, as explained in Section~\ref{sec:ward}, all states with $E>0$ organize into supermultiplets with representations $W\otimes\mathbf{256}$, where the supercharges act trivially on $W$, and for the two-graviton scattering states $W=\oplus_{m\geq0}(2m,0,0,0)$. The two-graviton representations $W\otimes\mathbf{256}$ further decompose into the irreps,
\begin{equation}
\begin{gathered}
(2m,0,0,0) \ ,\quad(n,1,0,0) \ ,\quad(2m+1,1,1,0) \ ,\\
(2n,1,1,1) \ ,\quad(2n,2,0,0) \ ,\\
(j,\tfrac12,\tfrac12,\tfrac12) \ ,\quad(j,\tfrac32,\tfrac12,\tfrac12) \ ,
\end{gathered}
\label{eq:reach}
\end{equation}
with $m\ge0$, $n\ge1$ and half-integer $j$, where $j\ge\tfrac32$ in the last family.

Consider now a state in a $Spin(9)$ representation that is not among the irreps~\eqref{eq:reach}. This state cannot be a scattering state of two gravitons. Intuitively, along the valley the strings stretched between the D0-branes cannot all be in their ground state, since the state would then be one of two gravitons. Because an excited string costs an energy that grows with the separation, it is possible for the two branes to be confined in such a sector, giving rise to normalizable states with $E>0$. A proof of this expectation has been proposed for the sectors with sufficiently large second row in their representation~\cite{OpenAIpos}, based on a mechanism that is essentially identical to the one we are discussing here.

Because of supersymmetry, we can also expect additional bound states. To be concrete, imagine we now have a bound state $|\psi\rangle$ that is not in any of the scattering irreps~\eqref{eq:reach}, but in whose supermultiplet there exists a state $|\tilde\psi\rangle=Q_\alpha|\psi\rangle$ in one of the scattering irreps~\eqref{eq:reach}. Then, $\|Q_\alpha\psi\|^2=E\|\psi\|^2$ implies that $|\tilde\psi\rangle$ is also normalizable, and it does not decay, since the scattering states of the same energy belong to multiplets with a different $W$. In other words, supersymmetry allows also for bound states in representations with a continuum of scattering states.\footnote{The same mechanism was noted in the model with four supercharges~\cite{Wosiek02}.}

The supermultiplets that are bound in this way are all those with $W$ other than $(2m,0,0,0)$. To see this, let $\lambda$ be the highest weight of such a $W$. The product $W\otimes\rep{84}$ contains the representation of highest weight $\lambda+(1,1,1,0)$, which is not in the list~\eqref{eq:reach}, so that every such multiplet has a member that can only occur as a bound state. For example, a bound state of the vector sector $\rep9$ has partners embedded in the continuum whatever its multiplet, since each of the ten multiplets $W\otimes\rep{256}$ that contain the $\rep9$ also contains sectors of eq.~\eqref{eq:reach}. For $W=\rep9$ these are the $\rep{16}$, $\rep{36}$, $\rep{128}$, $\rep{231}$, $\rep{432}$ and $\rep{576}$. Excitations of the stretched strings with $W=(2m,0,0,0)$, by contrast, have no such protection and can decay into the continuum.

For higher values of $N$, the above argument does not generalize in a straightforward way, since we have to account for the $N-1$ relative momenta, the $16(N-1)$ relative fermions, and the channels in which the D0-branes form bound clusters. Nonetheless, we would like to make the following observation. In the channel where all $N$ D0-branes are separated, the orbital wavefunction is a function of the $N-1$ relative momenta, \textit{i.e.} a sum of products of $N-1$ harmonics with a radial function (which is a singlet). For $N\geq 3$, we do not expect the exchange rules for the $N$ D0-branes to remove representations. The relative fermions contribute $\rep{256}^{\otimes N-1}$. Overall, we expect the scattering representations in this channel to take the form
\begin{equation}
\bigoplus_{\ell_i}\otimes_{i=1}^{N-1}(\ell_i,0,0,0)\otimes\rep{256}^{\otimes N-1} \ .
\end{equation}
For $N\geq 5$, representation theory tells us that every unitary irrep of $Spin(9)$ is contained inside $\otimes_{i=1}^{N-1}(\ell_i,0,0,0)\otimes\rep{256}^{\otimes N-1}$, since the harmonics of four momenta carry every tensor representation and the polarizations supply the spinors. This would suggest, in particular, that for $N\geq5$ the mechanism described above for protecting excited bound states from decaying does not apply, since the scattering states cover every possible representation. It would be interesting to test this expectation. For the intermediate cases with $N=3,4$ there may still be bound states with positive energy.

%% file: Sections/5.Outlook.tex
\section{Outlook}\label{sec:outlook}

In this article, we determined numerically the zero-energy bound state of the $SU(2)$ BFSS matrix model. Leveraging the fact that the state is a $Spin(9)\times SU(2)$ singlet, which results in a significant reduction of the underlying PDE problem, we obtained the wavefunction by implementing the variational method. Our solution has an overlap of $99.8\%$ with the true ground state. The solution also accurately reproduces the expected power-law decay of the state's tail, with its coefficient matching the value of the three-graviton coupling, while the moments of the state appear to be in very good agreement with the matrix bootstrap.

In Section~\ref{sec:ward} we showed that, for two D0-branes, the tail coefficient is fixed by the non-renormalized interaction of two distant D0-branes. It would be interesting to generalize this argument to the three-graviton coupling with general longitudinal momenta, for which the corresponding tails follow from the index computation of ref.~\cite{HM3}, see eq.~\eqref{eq:Chmgen}. In that case, however, comparing two phase shifts is no longer enough, since the D0-branes can then break into more than two clusters. It may be possible to present the argument directly in terms of the scattering amplitude, which could extend it to these tails as well.

In Section~\ref{sec:excited}, we discussed how supersymmetry and $Spin(9)$ representation theory allow for positive-energy bound states in the $SU(2)$ model, on top of the continuum of scattering states. An important question is to what extent these excited bound states persist for higher values of $N$. In this direction, we argued that their representation-theoretic protection mechanism fails for $N\geq 5$. Whether the $SU(N)$ model has excited bound states for $N\geq 5$ is then an outstanding question. These bound states are heavy, far from the low-energy regime of the matrix model that corresponds to M-theory, and \textit{a priori} they therefore do not appear to be directly relevant to the BFSS conjecture. Nonetheless, states of this kind may survive as metastable resonances at large $N$. It would be interesting to determine these resonances and to examine whether some of them behave as microstates of the black holes that the model describes in the 't Hooft and M-theory limits~\cite{BFKS,HorowitzMartinec,BFKS2,IMSY,Hanada13,Berkowitz16}. Finally, it would be desirable to extend these numerical methods to higher $N$, where the physics of the system is expected to be much richer, with $SU(3)$ BFSS being the natural next target.

%% file: Sections/S.Supplemental.tex
\section{Conventions}\label{app:conventions}

In the conventions used in this article, the BFSS Hamiltonian~\eqref{eq:H} reads
\begin{equation}\label{eq:H_app}
H=\Tr\Big(\tfrac12 \big(P^{I}\big)^2-\tfrac{g^2}4[X^I,X^J]^2-\tfrac g2\psi^T\gamma^I[X^I,\psi]\Big) \ ,
\end{equation}
where $X$ and $\psi$ are a $Spin(9)$ vector and spinor, respectively, and $P$ is the momentum conjugate to $X$, all in the adjoint representation of the $SU(N)$ gauge group. The $Spin(9)$ gamma matrices are the $16\times 16$ real symmetric matrices $\gamma^I$ with $\{\gamma^I,\gamma^J\}=2\delta^{IJ}$, while $g^2$ is the Yang--Mills coupling with dimensions of energy cubed. We denote by capital Latin letters $I,J=1,\dots,9$ the vector and $A,B=1,\dots,N^2-1$ the gauge indices, and by lowercase Greek letters $\alpha=1,\dots,16$ the spinor indices. Upon quantization, the fields in the Hamiltonian~\eqref{eq:H_app} obey the (anti)commutation relations,
\begin{equation}
\big[X^{IA},P^{JB}\big]=i\delta^{IJ}\delta^{AB} \ , \quad \big\lbrace \psi^A_\alpha,\psi^B_\beta\big\rbrace=\delta_{\alpha\beta}\delta^{AB} \ ,
\end{equation}
where the components are defined by $X^I=X^{IA}T^A$, and similarly for $P^I$ and $\psi_\alpha$, in a basis of $SU(N)$ generators with $[T^A,T^B]=if^{ABC}T^C$ and $\Tr(T^AT^B)=\delta^{AB}$. In the above basis, the sixteen supercharges take the form,
\begin{equation}\label{eq:Q_app}
Q_\alpha=P^{IA}(\gamma^I\psi^A)_\alpha+\frac g 2 f^{ABC}X^{IA}X^{JB}(\gamma^{IJ}\psi^C)_\alpha \ ,
\end{equation}
for $\gamma^{IJ}:=\frac 12 [\gamma^I,\gamma^J]$, with the corresponding algebra being
\begin{equation}\label{eq:SUSY_algebra}
\big\lbrace Q_\alpha,Q_\beta\big\rbrace=2\delta_{\alpha\beta}H+2g\gamma^I_{\alpha\beta}X^{IA}G^A \ , \quad G^A=f^{ABC}\big(X^{IB}P^{IC}-\frac i2 \psi^B_\alpha\psi^C_\alpha\big) \ ,
\end{equation}
where $G^A$ generates the $SU(N)$ gauge transformations and annihilates the physical states, which is the Gauss constraint. In other words, the supersymmetry algebra closes only on the physical $SU(N)$-singlet states. The $Spin(9)$ generators read,
\begin{equation}\label{eq:J_IJ}
J^{IJ}=X^{IA}P^{JA}-X^{JA}P^{IA}-\frac i 4 \psi^A_\alpha\gamma^{IJ}_{\alpha\beta}\psi^A_\beta \ .
\end{equation}
The generators $G^A$ and $J^{IJ}$ act on the adjoint fields $\Phi^A\in\lbrace X^{IA},\psi^A_\alpha\rbrace$ as
\begin{equation}\label{eq:G_J_indep_rotations}
\big[G^A,\Phi^D\big]=-if^{ABD}\Phi^B \ , \quad \big[J^{IJ},X^{KA}\big]=-i\big(\delta^{JK}X^{IA}-\delta^{IK}X^{JA}\big) \ , \quad \big[J^{IJ},\psi^A\big]=\frac i 2 \gamma^{IJ}\psi^A \ .
\end{equation}
An important observation that we will use later is that both the $SU(N)$ and the $Spin(9)$ generators are sums of a bosonic and a fermionic part, which rotate the bosons and the fermions independently.

For the specific case of $SU(2)$ BFSS considered here, we follow the conventions of ref.~\cite{FGHHY}. In particular, we label the bosonic coordinates as $q_{IA}$, the momenta $p_{IA}=-i\,\partial/\partial q_{IA}$, and the fermions as $\Theta_{\alpha A}$, with $\{\Theta_{\alpha A},\Theta_{\beta B}\}=\delta_{\alpha\beta}\delta_{AB}$, and we write $\vec q_I$, $\vec p_I$ and $\vec\Theta_\alpha$ for the corresponding vectors in color space. The Hamiltonian then is
\begin{equation}
H=\sum_{I}\vec p_{I}\cdot\vec p_I +\sum_{I<J}|\vec q_I\times\vec q_J|^2+i\sum_{I,\alpha,\beta}\vec q_I\cdot(\vec\Theta_\alpha\times\vec\Theta_\beta)\,\gamma^I_{\alpha\beta} \ ,
\label{eq:Hfghhy}
\end{equation}
where we reinstated the sums. To relate it to the general expression~\eqref{eq:H_app}, we expand the matrices in terms of the $SU(2)$ generators $t^A=\sigma^A/2$, as
\begin{equation}
X^I=q_{IA}\,t^A \ , \qquad P^I=2\,p_{IA}\,t^A \ , \qquad \psi_\alpha=\sqrt2\,\Theta_{\alpha A}\,t^A \ ,
\end{equation}
in which case the Hamiltonian~\eqref{eq:Hfghhy} coincides with eq.~\eqref{eq:H_app} at $g=2$. Note that the generators normalized as in the first part of this Appendix are $T^A=\sqrt2\,t^A$, and therefore the above identification amounts to $X^{IA}=q_{IA}/\sqrt2$, $P^{IA}=\sqrt2\,p_{IA}$, $\psi^A_\alpha=\Theta_{\alpha A}$ and $f^{ABC}=\sqrt2\,\epsilon_{ABC}$.

In these variables the supercharges~\eqref{eq:Q_app} take the form
\begin{equation}
Q_\alpha=\sqrt2\,\Big[\sum_{I,\beta}\gamma^I_{\alpha\beta}\,\vec p_I\cdot\vec\Theta_\beta+\sum_{I<J}\sum_{\beta}\gamma^{IJ}_{\alpha\beta}\,(\vec q_I\times\vec q_J)\cdot\vec\Theta_\beta\Big] \ ,
\end{equation}
and they obey $\{Q_\alpha,Q_\beta\}=2\delta_{\alpha\beta}H$ on gauge-invariant states, with $H$ the Hamiltonian~\eqref{eq:Hfghhy}.

Let us finally comment on the physical meaning of these coordinates. The potential in eq.~\eqref{eq:Hfghhy} vanishes along the valley of commuting matrices, which up to a gauge rotation is parametrized as $q_{IA}=R\,\hat r_I\,\delta_{A3}$, with $|\hat r|=1$. There, the matrices $X^I$ have eigenvalues $\pm\tfrac12R\,\hat r_I$, so that the coordinate $R$ is the separation between the two D0-branes.

As for the units, throughout the paper we set $g=2$ and the radius of the lightlike circle to $R_-=1$, in which case each D0-brane has unit mass and the eleven-dimensional Planck length, in the conventions of ref.~\cite{HM3}, is $\lp=(16\pi^2)^{-1/6}\simeq0.430$, with $\kappa_{11}^2=16\pi G_{11}=(2\pi)^8\lp^9$.

\section{Details on the Value of the Three-Graviton Coupling}
\label{ap:three_grav_coupling}

The Hamiltonian~\eqref{eq:H} with $g=2$ coincides with the conventions of ref.~\cite{HM3} for the radius of the lightlike circle $R_-=1$ and $\lp^6=1/(16\pi^2)$. For $N_1=N_2=1$, and stripped of the factor $(2\pi)^9$ and of the delta functions of energy, transverse momentum and Grassmann variables, their amplitude reads
\begin{equation}
A=8\sqrt2(2\pi)^3\lp^{9/2}R_-\frac{p_9^2}{\sqrt{R_9}} \ , \qquad p_9=\frac{n_1-n_2}{2R_9} \ ,
\end{equation}
where $p_9$ is the relative momentum along the circle, conjugate to the separation. In ref.~\cite{HM3} one transverse direction, $x^9$, is also compactified on a circle of radius $R_9$, and the two states carry momenta $n_1/R_9$ and $n_2/R_9$ along it. The delta function of the relative Grassmann variables selects the relative polarization $\varepsilon_I\varepsilon_J\ket{h_{IJ}}$, with $\varepsilon=(e_9+ie_8)/\sqrt2$, which has unit norm. The comparison with eq.~\eqref{eq:split-vertex} requires three steps.

The first step concerns the normalization of the states. The amplitude refers to states of unit norm on a circle of length $2\pi R_9$, and such a state equals $1/\sqrt{2\pi R_9}$ times a continuum state normalized as $\langle k'|k\rangle=(2\pi)^9\delta^9(k'-k)$. The three external states therefore contribute a factor $(2\pi R_9)^{3/2}$, while the conversion of the Kronecker delta of momentum conservation along the circle into a Dirac delta function removes one factor $2\pi R_9$. Between continuum states, with the center of mass removed and $\langle B|B\rangle=1$, the vertex at $R_-=1$ is therefore
\begin{equation}
A_9=\sqrt{2\pi R_9}A=8\sqrt2(2\pi)^{7/2}\lp^{9/2}p_9^2 \ .
\end{equation}
The second step has to do with the polarization. More specifically, the amplitude is evaluated at the complex null relative momentum $k=p_9(e_9+ie_8)=\sqrt2p_9\varepsilon$, for which $k_Ik_J\ket{h_{IJ}}=2p_9^2\varepsilon_I\varepsilon_J\ket{h_{IJ}}$ has norm $2p_9^2$ in the $|h_{IJ}\rangle$ normalization presented after eq.~\eqref{eq:tail}. Lastly, the third step concerns the exchange of the two D0-branes. By the exchange rule discussion before eq.~\eqref{eq:exchange}, the gauge-invariant two-graviton states are $(\ket{k}\otimes\ket{a}+\ket{-k}\otimes(-1)^F\ket{a})/\sqrt2$, where $(-1)^F$ acts on the relative polarization and $k$ is restricted to a half-space. Since $\ket{B}$ is invariant under the exchange, its vertex with these states is $\sqrt2$ times $\bra{k,a}V\ket{B}$. Combining the three steps yields
\begin{equation}
\sqrt2\frac{\Omega_8}{9}|C|\cdot2p_9^2=A_9 \ , \qquad |C|=\frac{945\sqrt2}{8\pi^2} \ ,
\label{eq:match}
\end{equation}
and consequently, by virtue of eq.~\eqref{eq:PR},
\begin{equation}
|C|^2=\frac{893025}{32\pi^4}\simeq286.493 \ , \qquad P(R)\simeq\frac{7560}{R^{10}} \ .
\end{equation}

The same conversion also applies to the general split of $N_3$ units of longitudinal momentum into $N_1$ and $N_2$. In that case the relative energy is $k^2/(2m)$, with $m=N_1N_2/N_3$, instead of $k^2$, and the factor associated with the exchange is present only for $N_1=N_2$. Summing over the ways of distributing the D0-branes between the two clusters, the density is
\begin{equation}
P_{N_1,N_2}(R)dR=\frac{945}{2(1+\delta_{N_1N_2})}\frac{N_3^5}{N_1^3N_2^3}\frac{dR}{R^{10}} \ .
\label{eq:Chmgen}
\end{equation}

\section{Reduction for the Ground State of \texorpdfstring{$SU(2)$}{SU(2)} BFSS}\label{app:reduction}

After removing the free center of mass, the $SU(2)$ problem has $9\times(2^2-1)=27$ bosonic coordinates $q_{IA}$, which we arrange into a real $9\times3$ matrix $q$, and $16\times(2^2-1)=48$ real fermions $\Theta_{\alpha A}$, which combine into $24$ complex fermion modes. A general state can be expanded as
\begin{equation}
|\Psi\rangle=\sum_{s=1}^{2^{24}}\Psi_s(q)\,|s\rangle \ ,
\end{equation}
where $|s\rangle$ runs over a basis of the $2^{24}$-dimensional fermionic Fock space and the coefficients $\Psi_s$ are functions of the $27$ bosonic variables. Since $H$ preserves fermion parity, only $2^{23}$ of these functions are nonzero for an eigenstate of definite fermion parity. Physical states must in addition obey the Gauss constraint $G^A|\Psi\rangle=0$.

The Witten index computations guarantee the existence of a normalizable zero-energy state~\cite{Yi,SethiStern}, and every zero-energy state is a $Spin(9)$ singlet~\cite{SethiStern2,HaslerHoppe02}. Singlets are bosonic, since $(-1)^F$ is the rotation by $2\pi$ in $Spin(9)$, so the number of zero-energy states equals the index, which is one, and the ground state is unique. It is annihilated by both $G^A$ and $J^{IJ}$. The combined group $\mathcal G=Spin(9)\times SU(2)$ acts on the bosonic coordinates through $SO(9)\times SO(3)$, as
\begin{equation}
q\;\to\;U\,q\,V^T \ , \qquad U\in SO(9)\ ,\quad V\in SO(3) \ .
\end{equation}
By eq.~\eqref{eq:G_J_indep_rotations}, both $G^A$ and $J^{IJ}$ are sums of a bosonic part, which generates this transformation of $q$, and a fermionic part, which acts only on the Fock space. Therefore, invariance of $|\Psi\rangle$ under $\mathcal G$ is equivalent to
\begin{equation}\label{eq:equivariance}
|\Psi(UqV^T)\rangle=D_F(U,V)\,|\Psi(q)\rangle \ ,
\end{equation}
where $|\Psi(q)\rangle=\sum_s\Psi_s(q)|s\rangle$ is the Fock-space vector at the point $q$ and $D_F(U,V)$ is the unitary operator on the Fock space that implements the lift of $(U,V)$ to $Spin(9)\times SU(2)$. It is generated by the fermionic parts $-\tfrac i4\psi^A\gamma^{IJ}\psi^A$ of $J^{IJ}$ and $-\tfrac i2f^{ABC}\psi^B_\alpha\psi^C_\alpha$ of $G^A$. Equation~\eqref{eq:equivariance} relates the values of the wavefunction at all points of a $\mathcal G$-orbit.

\subsection{Bosonic reduction}

By the singular value decomposition, every $9\times3$ matrix can be written as $q=U\Sigma V^T$ with $U\in SO(9)$, $V\in SO(3)$ and
\begin{equation}
\Sigma=\begin{pmatrix}\mathrm{diag}(r_1,r_2,r_3)\\ 0_{6\times3}\end{pmatrix} \ , \qquad r_1\ge r_2\ge r_3\ge0 \ .
\end{equation}
Here $r_i^2$ are the eigenvalues of the color Gram matrix $q^Tq$, which is invariant under $U$ and transforms by conjugation under $V$, so the $r_i$ are invariant under $\mathcal G$. With $U$ and $V$ restricted to rotations, the $r_i$ can still be taken non-negative, because the six directions orthogonal to the image of $q$ allow any sign flip to be completed to a rotation in $SO(9)$. By eq.~\eqref{eq:equivariance}, the ground state is completely determined by its values $|\Psi(\Sigma)\rangle$ on this three-dimensional slice. Changing variables from $q$ to $(r_i,U,V)$ gives, up to a constant, the measure
\begin{equation}\label{eq:Jacobian}
d^{27}q\propto J(r)\,d^3r\,dU\,dV \ ,\qquad J(r)=\prod_{i<j}|r_i^2-r_j^2|\prod_ir_i^{6} \ ,
\end{equation}
with $dU$ and $dV$ the Haar measures.

\paragraph{Stabilizer.} If $(U,V)$ leaves $\Sigma$ invariant, \textit{i.e.} $U\Sigma V^T=\Sigma$, eq.~\eqref{eq:equivariance} gives
\begin{equation}\label{eq:stabilizer-condition}
|\Psi(\Sigma)\rangle=D_F(U,V)\,|\Psi(\Sigma)\rangle \ ,
\end{equation}
so the fermionic state on the slice must be invariant under the stabilizer of $\Sigma$. For generic $\Sigma$, with distinct and nonzero $r_i$, the condition $U\Sigma=\Sigma V$ forces $U$ to be block diagonal, $U=u\oplus R_6$, and $u\,\mathrm{diag}(r_i)=\mathrm{diag}(r_i)\,V$ forces $u=V=\mathrm{diag}(\epsilon_1,\epsilon_2,\epsilon_3)$ with $\epsilon_i=\pm1$. Requiring $\det V=\det U=1$ then gives
\begin{equation}
\mathrm{Stab}(\Sigma)=\big\{(\mathrm{diag}(\epsilon)\oplus R_6,\, \mathrm{diag}(\epsilon)):\ \epsilon_1\epsilon_2\epsilon_3=1,\ R_6\in SO(6)\big\}\simeq SO(6)\times\mathbb Z_2\times\mathbb Z_2 \ .
\end{equation}
The $SO(6)$ rotates the six directions orthogonal to the image of $q$, and each non-trivial element of $\mathbb Z_2\times\mathbb Z_2$ flips the signs of two color axes, compensated by a $\pi$ rotation in the corresponding plane of $\mathbb R^9$. Since $\mathcal G$ has dimension $36+3=39$ and the stabilizer has dimension $15$, generic orbits are $24$-dimensional, leaving $27-24=3$ bosonic coordinates, $r_1,r_2,r_3$.

\subsection{Fermionic reduction}

Let us now discuss what happens on the fermion side. After choosing the slice for the bosonic matrices in which $q=\Sigma$, eq.~\eqref{eq:equivariance} requires the fermionic states to be invariant under all $(U,V)$ in the stabilizer $\mathrm{Stab}(\Sigma)$. The stabilizer acts on the first three axes of $\mathbb R^9$ through the $\mathbb Z_2\times\mathbb Z_2$ and on the last six through $SO(6)$. It is therefore convenient to decompose
\begin{equation}
Spin(9)\supset Spin(3)\times Spin(6)\simeq SU(2)_R\times SU(4) \ ,
\end{equation}
where the product denotes the covering group (the embedded subgroup is its quotient by the diagonal $\mathbb Z_2$), $SU(2)_R$ is the spin cover of rotations of $e_1,e_2,e_3$, and $SU(4)$ that of rotations of $e_4,\dots,e_9$, where $e_I$ is a basis for $\mathbb R^9$. Under this subgroup the real $\mathbf{16}$ of $Spin(9)$ becomes $(\mathbf 2,\mathbf 4)\oplus(\mathbf 2,\bar{\mathbf 4})$ after complexification, with the reality condition relating the two terms. Hence the sixteen real components $\Theta_\alpha$ can be packaged into eight complex combinations transforming in $(\mathbf 2,\mathbf 4)$. Including the color index $A=1,2,3$, the $48$ real fermions $\Theta_{\alpha A}$ become $24$ complex annihilation operators
\begin{equation}
c_{m f} \ ,\qquad m=1,\dots,4 \ ,\qquad f=(a,A)\in\{1,2\}\times\{1,2,3\} \ ,
\end{equation}
where $m$ is an $SU(4)$ index and the flavor index $f$ combines the $SU(2)_R$ index $a$ with the color index $A$. Writing $Y:=\mathbb C^2\otimes\mathbb C^3\simeq\mathbb C^6$, the one-fermion states span $\mathbb C^4\otimes Y$, and the Fock space is its exterior algebra,
\begin{equation}
\mathcal F=\Lambda^\bullet\big(\mathbb C^4\otimes Y\big) \ , \qquad \dim\mathcal F=2^{24} \ .
\end{equation}
$SU(4)$ acts only on the first factor, while the $\mathbb Z_2\times\mathbb Z_2$ acts only on the second. We will now successively restrict to $SU(4)$ and $\mathbb Z_2\times\mathbb Z_2$ invariant states.

\paragraph{$SU(4)$ singlets.} The decomposition of $\mathcal F$ under $GL(4)\times GL(6)$ is given by skew Howe duality,
\begin{equation}
\Lambda^\bullet\big(\mathbb C^4\otimes Y\big)=\bigoplus_{\lambda}S_\lambda(\mathbb C^4)\otimes S_{\lambda'}(Y) \ ,
\label{eq:skewHowe}
\end{equation}
where $\lambda$ runs over Young diagrams with at most four rows and six columns, $\lambda'$ denotes the transposed diagram, and $S_\lambda$ is the corresponding Schur functor. This is simply the statement that antisymmetry under exchange of two fermions requires the $SU(4)$ and flavor labels to come in opposite symmetry representations. In particular, the factor $S_\lambda(\mathbb C^4)$ is an $SU(4)$ singlet if and only if $\lambda$ consists of complete columns of height four, $\lambda=(k,k,k,k),\,k=0,1,\dots,6$, each column corresponding to an $\epsilon^{mnpq}$-symbol contraction. The $SU(4)$-invariant subspace is therefore
\begin{equation}
\mathcal F^{SU(4)}=\bigoplus_{k=0}^{6}S_{(4^k)}(Y) \ ,
\end{equation}
where $(4^k)$ denotes the diagram with $k$ rows of length four. The sector labeled by $k$ has fermion number $4k$ and is spanned by products of $k$ baryon operators
\begin{equation}
B_{f_1f_2f_3f_4}=\epsilon^{mnpq}c^\dagger_{mf_1}c^\dagger_{nf_2}c^\dagger_{pf_3}c^\dagger_{qf_4} \ ,
\end{equation}
acting on the Fock vacuum. In particular, all $SU(4)$-invariant states have even fermion number, consistent with a bosonic ground state. The dimensions $d_k:=\dim S_{(4^k)}(\mathbb C^6)$ follow from the hook-content formula,
\begin{equation}
\dim S_\mu(\mathbb C^n)=\prod_{(i,j)\in\mu}\frac{n+j-i}{h(i,j)} \ ,
\end{equation}
where $h(i,j)$ is the hook length of the box in row $i$ and column $j$. For example, $d_1=\dim S_{(4)}(\mathbb C^6)=\binom94=126$, corresponding to the totally symmetric flavor content of a single baryon. The full list is given in Table~\ref{tab:fermion-count}. The above implies that the total number of $SU(4)$ singlet states in $\mathcal F$ is given by $\sum_kd_k=7898$.

\paragraph{$\mathbb Z_2\times\mathbb Z_2$ projection.} The remaining stabilizer elements commute with $SU(4)$ and therefore act within each $S_{(4^k)}(Y)$. Consider the element that flips the signs of color axes $1$ and $2$, compensated by a $\pi$ rotation in the $(1,2)$ plane of $\mathbb R^9$. The $\pi$ rotation acts on the $SU(2)_R$ doublet as $\pm i\sigma_3$, and the flip acts on the color index as $\mathrm{diag}(-1,-1,1)$. The element therefore acts diagonally on $Y$, with eigenvalues that in a suitable ordering of the basis are
\begin{equation}
x=(\,i,\,i,\,i,\,-i,\,-i,\,-i\,) \ ,
\end{equation}
and the two other non-trivial elements are obtained by relabeling the axes and have the same eigenvalues. The trace of each of them on $S_{(4^k)}(Y)$ is the Schur polynomial $s_{(4^k)}(x)$. Since $s_\mu$ is homogeneous of degree $|\mu|=4k$, the overall factors of $i$ drop out, and one gets
\begin{equation}
\chi_k := s_{(4^k)}(i,i,i,-i,-i,-i)=s_{(4^k)}(1,1,1,-1,-1,-1) \ .
\end{equation}
Equivalently, $\chi_k$ counts the semistandard Young tableaux of shape $(4^k)$ with entries in $\{1,\dots,6\}$, weighted by $(-1)^n$, where $n$ is the number of entries in $\{4,5,6\}$. For $k=1$ this gives
\begin{equation}
\chi_1=\sum_{n=0}^{4}(-1)^n\binom{n+2}{2}\binom{6-n}{2}=6 \ .
\end{equation}
Then, the number of states invariant under $\mathbb Z_2\times\mathbb Z_2$ in each sector is the average of the trace over the four elements of the group,
\begin{equation}
n_k=\tfrac14\big(d_k+3\chi_k\big) \ .
\end{equation}
The ambiguity in lifting each element to $Spin(9)$ is a factor of $(-1)^F$, which acts trivially on these even states. The number of $\mathbb Z_2\times \mathbb Z_2$ singlets inside the $SU(4)$ singlets is given at each level $k$ in Table~\ref{tab:fermion-count}, yielding a total of $2066$ remaining states.

\par\noindent\begin{minipage}{\linewidth}
\begin{center}
\small\setlength{\tabcolsep}{8pt}
\begin{tabular}{@{}lrrrrrrrr@{}}
\toprule
$k$ & 0 & 1 & 2 & 3 & 4 & 5 & 6 & total\\
\midrule
fermion number $4k$ & 0 & 4 & 8 & 12 & 16 & 20 & 24 & \\
$d_k=\dim S_{(4^k)}(\mathbb C^6)$ & 1 & 126 & 1764 & 4116 & 1764 & 126 & 1 & 7898\\
$\chi_k$ & 1 & 6 & 36 & 36 & 36 & 6 & 1 & \\
$n_k=\tfrac14(d_k+3\chi_k)$ & 1 & 36 & 468 & 1056 & 468 & 36 & 1 & 2066\\
\bottomrule
\end{tabular}
\end{center}
{\small\refstepcounter{table}\label{tab:fermion-count}\noindent TABLE~\thetable.\ Counting of fermionic states on the slice. $d_k$ is the number of $SU(4)$ singlets with fermion number $4k$, $\chi_k$ the trace of a non-trivial element of $\mathbb Z_2\times\mathbb Z_2$ on them, and $n_k$ the number of states invariant under the full stabilizer.\par}
\end{minipage}
\medskip

In conclusion, at each point $(r_1,r_2,r_3)$ of the slice the fermionic part of the ground state lies in a fixed $2066$-dimensional space,
\begin{equation}
\ket{\Psi(\Sigma)}\in\mathcal V :=\mathcal F^{SU(4)\times\mathbb Z_2\times\mathbb Z_2} \ , \qquad \dim \mathcal V =2066 \ ,
\end{equation}
and the ground-state problem reduces to $2066$ coupled PDEs for functions of three variables $\Psi_s(r_1,r_2,r_3)$.

\subsection{The reduced Hamiltonian}

We now derive the action of the Hamiltonian~\eqref{eq:Hfghhy} on the $Spin(9)\times SU(2)$-singlet ground state $|\Psi(\Sigma)\rangle\in\mathcal V$. Throughout this subsection, we use $i,j,k\in\{1,2,3\}$ and $\nu\in\{4,\dots,9\}$.

\paragraph{Kinetic term.} The kinetic term in eq.~\eqref{eq:Hfghhy} is minus the Laplacian of the flat metric $ds^2=\sum_{I,A}dq_{IA}^2=\Tr(dq^Tdq)$ on $\mathbb R^{27}$. Near a generic point we use the coordinates $q=U\Sigma V^T$, and we describe infinitesimal changes of $U$ and $V$ by the antisymmetric matrices
\begin{equation}
\omega=U^TdU\in so(9) \ , \qquad \theta=V^TdV\in so(3) \ .
\end{equation}
Differentiating gives $dq=U(d\Sigma+\omega\Sigma-\Sigma\theta)V^T$. Since $U,V$ are orthogonal, we have
\begin{equation}
ds^2=\Tr\big(M^TM\big) \ , \qquad M\equiv d\Sigma+\omega\Sigma-\Sigma\theta \ .
\end{equation}
Using $\Sigma_{IA}=r_A\delta_{IA}$ and the antisymmetry of $\omega$ and $\theta$, the various entries of the $9\times3$ matrix $M$ read,
\begin{equation}
M_{ii}=dr_i \ , \qquad M_{ij}=\omega_{ij}r_j-r_i\theta_{ij}\quad(i\neq j) \ , \qquad M_{\nu i}=\omega_{\nu i}r_i \ .
\end{equation}
The rotations $\omega_{\nu\nu'}$ do not appear, since they belong to the $SO(6)$ stabilizer. The off-diagonal entries couple $\omega_{ij}$ and $\theta_{ij}$. Defining $\omega_{ij}=\omega^+_{ij}+\omega^-_{ij}$ and $\theta_{ij}=\omega^+_{ij}-\omega^-_{ij}$ removes the cross terms, and the metric becomes diagonal,
\begin{equation}\label{eq:slice-metric}
ds^2=\sum_idr_i^2+\sum_{i<j}\Big[2(r_i-r_j)^2\big(\omega^+_{ij}\big)^2+2(r_i+r_j)^2\big(\omega^-_{ij}\big)^2\Big]+\sum_{i,\nu}r_i^2\omega_{\nu i}^2 \ .
\end{equation}
The $27$ directions are thus the three slice coordinates $r_i$, together with $24$ directions parallel to the orbit. Up to a constant, the volume element of this metric is the Haar measure of $SO(9)\times SO(3)$ times $J(r)d^3r$, for $J(r)$ as in eq.~\eqref{eq:Jacobian}.

Let $E^\pm_{ij}$ and $E_{\nu i}$ be the vector fields dual to the one-forms $\omega^\pm_{ij}$ and $\omega_{\nu i}$ of eq.~\eqref{eq:slice-metric}, which span the $24$ orbit directions. The Laplacian then takes the form
\begin{equation}\label{eq:Laplacian-slice}
\Delta=\sum_i\frac1J\frac{\partial}{\partial r_i}\Big(J\frac{\partial}{\partial r_i}\Big)+\sum_{i<j}\left[\frac{\big(E^+_{ij}\big)^2}{2(r_i-r_j)^2}+\frac{\big(E^-_{ij}\big)^2}{2(r_i+r_j)^2}\right]+\sum_{i,\nu}\frac{E_{\nu i}^2}{r_i^2} \ .
\end{equation}
This follows from two observations. First, the metric~\eqref{eq:slice-metric} is diagonal with coefficients that depend only on $r$; second, these vector fields are divergence-free with respect to the Haar measure, since $SO(9)$ and $SO(3)$ are unimodular. The next step is to express the derivatives along the orbits in terms of certain operators, as determined by equivariance. These operators are built from the fermionic parts of the $Spin(9)$ generators $J^{IJ}$~\eqref{eq:J_IJ} and of the color generators $G^k/\sqrt2$~\eqref{eq:SUSY_algebra}, namely
\begin{equation}
S^{IJ}=-\frac{i}{4}\Theta_{\alpha A}\gamma^{IJ}_{\alpha\beta}\Theta_{\beta A} \ , \qquad S^k=-\frac{i}{2}\epsilon_{kBC}\Theta_{\alpha B}\Theta_{\alpha C} \ .
\end{equation}
They rotate the spinor and color indices of the fermions, respectively, exactly as the orbital parts of $J^{IJ}$ and $G^k$ rotate the bosons. More specifically, the direction $\omega_{\nu i}$ is a spatial rotation in the $(i,\nu)$ plane, with generator $S^{i\nu}$. The directions $\omega^+_{ij}$ and $\omega^-_{ij}$ combine a spatial rotation in the $(i,j)$ plane with a color rotation about the axis $k$, with generators
\begin{equation}\label{eq:red1}
K_{S,ij}=S^{ij}+\epsilon_{ijk}S^k \ , \qquad K_{D,ij}=S^{ij}-\epsilon_{ijk}S^k \ .
\end{equation}
The relative sign is fixed by requiring that $K_{S,ij}$ generates the rotation that leaves $\Sigma$ invariant when $r_i=r_j$. Since the total generators annihilate the state, a derivative along the orbit acts on it as $\pm i$ times the corresponding fermionic generator, so that $\big(E^+_{ij}\big)^2$, $\big(E^-_{ij}\big)^2$ and $E_{\nu i}^2$ act as $-K_{S,ij}^2$, $-K_{D,ij}^2$ and $-\big(S^{i\nu}\big)^2$. Evaluating the kinetic term on the slice, we then find
\begin{equation}\label{eq:red2}
\sum_{I,A}p_{IA}^2\Big|_{\Sigma}=-\sum_i\frac1J\frac{\partial}{\partial r_i}\Big(J\frac{\partial}{\partial r_i}\Big)+\sum_{i<j}\left[\frac{K_{S,ij}^2}{2(r_i-r_j)^2}+\frac{K_{D,ij}^2}{2(r_i+r_j)^2}\right]+\sum_i\frac{O_i}{r_i^2} \ , \qquad O_i=\sum_{\nu=4}^9\big(S^{i\nu}\big)^2 \ .
\end{equation}

\paragraph{Potential and Yukawa terms.} The interaction terms contain no derivatives and can be evaluated directly on the slice $\Sigma$, where $\vec q_i=r_i\vec e_i$, with $\vec e_i$ the unit vectors of color space, and $\vec q_\nu=0$. The bosonic potential, then, gives
\begin{equation}
\sum_{I<J}\big|\vec q_I\times\vec q_J\big|^2\Big|_{q=\Sigma}=\sum_{i<j}r_i^2r_j^2 \ ,
\end{equation}
and only the $I=1,2,3$ components contribute to the Yukawa term,
\begin{equation}\label{eq:red3}
i\sum_I\sum_{\alpha,\beta}\gamma^I_{\alpha\beta}\vec q_I\cdot\big(\vec\Theta_\alpha\times\vec\Theta_\beta\big)\Big|_{q=\Sigma}=\sum_ir_iY_i \ , \qquad Y_i=i\sum_{\alpha,\beta}\sum_{B,C}\epsilon_{iBC}\gamma^i_{\alpha\beta}\Theta_{\alpha B}\Theta_{\beta C} \ ,
\end{equation}
with no sum over $i$ in the definition of $Y_i$.

\paragraph{Reduced Hamiltonian.} Since $H$ commutes with $Spin(9)\times SU(2)$, $H|\Psi\rangle$ obeys the same equivariance relation as $|\Psi\rangle$~\eqref{eq:equivariance}. This means that it vanishes everywhere if and only if it vanishes on the slice. Putting everything together, the reduced Hamiltonian on the slice $\Sigma$ becomes,
\begin{equation}\label{eq:Hred}
H_{\rm red}=-\sum_i\frac1J\frac{\partial}{\partial r_i}\Big(J\frac{\partial}{\partial r_i}\Big)+\sum_{i<j}\left[\frac{K_{S,ij}^2}{2(r_i-r_j)^2}+\frac{K_{D,ij}^2}{2(r_i+r_j)^2}\right]+\sum_i\frac{O_i}{r_i^2}+\sum_{i<j}r_i^2r_j^2+\sum_ir_iY_i \ .
\end{equation}
Every term in eq.~\eqref{eq:Hred} commutes with the stabilizer, so $H_{\rm red}$ maps $\mathcal V$ to itself. It is Hermitian with respect to the inner product
\begin{equation}\label{eq:inner}
\langle\Phi_1|\Phi_2\rangle=\int_{r_1\geq r_2\geq r_3\geq0}d^3r\,J(r)\,\Phi_1(r)^\dagger\Phi_2(r) \ ,
\end{equation}
which coincides with the inner product of the full wavefunctions up to the volume of the symmetry group. The coefficients of eq.~\eqref{eq:Hred} are singular on the walls $r_i=r_j$ and $r_i=0$, where the orbits degenerate and the stabilizer is enlarged. Regularity there requires the state to be invariant under the enlarged stabilizer, which provides the boundary conditions of the problem.

Note, lastly, that the above reduction is specific to the $SU(2)$ model. It relies on two properties that do not hold for $N\geq3$: the adjoint action of $SU(2)$ is the full rotation group $SO(3)$ of color space, so that $q$ can be brought to the form above, and these configurations have a continuous stabilizer, which reduces the fermionic space to $2066$ states.

\section{Details on the Numerical Solution}\label{app:numerics}

Here we describe how the reduced problem of Appendix~\ref{app:reduction} is solved numerically, and elaborate on how the coefficient of the tail and the moments of the wavefunction quoted in the main text are extracted from the solution.

\subsection{Solving for the ground state}

We solve for the normalizable zero-energy eigenstate of the reduced Hamiltonian~\eqref{eq:Hred_main-text} by using the variational method. In particular, we minimize the energy~\eqref{eq:form} over the trial states~\eqref{eq:ansatz}, in the box $0\leq a,b\leq3$, $0\leq c\leq L_c$. This requires specifying the mesh that defines the basis functions $N_n$, the fermionic channels $u_{n,m}$ and the solution of the resulting eigenvalue problem, which we describe in turn.

\paragraph{Mesh.} Across the valley, each of the intervals $0\leq a\leq3$ and $0\leq b\leq3$ is divided into $n_\perp=10$ or $12$ elements of lengths between $0.25$ and $0.36$, on which the $N_n$ are cubic or quartic polynomials. Along the valley direction $c$, the elements start at $0.15$ near $c=0$ and grow by a factor of $1.4$ up to $0.5$, which gives $32$ elements for $L_c=15$ and $46$ for $L_c=22$, or up to $0.25$ for our best state, which has $89$ elements, and the $N_n$ are quadratic in $c$. A polynomial of degree $q$ in a given direction is fixed by its values at $q+1$ equally spaced nodes of the element, and neighboring elements share the nodes on their common faces, so that the $N_n$ are continuous (as required by the first derivatives in eq.~\eqref{eq:form}). The nodes on the outer faces carry no basis function, so that the cubic $10\times10$ mesh with $L_c=15$ has $30\times30\times64=57600$ nodes. All integrals are computed with six Gauss points per direction in each element, except for the quartic states in the box $L_c=22$, including our best state, where we use sixteen points in $a$ and $b$. We keep a node only if the weighted norm $\int d^3rJ\eta^2N_n^2$ of its basis function exceeds a tolerance $\tau$ times the largest one. For instance, on the cubic $10\times10$ mesh, $\tau=10^{-12}$ keeps $18460$ nodes and $\tau=10^{-16}$ keeps $22643$, which lowers the variational energy $E_h$ from $0.0087$ to $0.0034$, while $\tau=10^{-24}$ changes it by only $10^{-5}$.

\paragraph{Fermionic channels.} The fermionic operators in eq.~\eqref{eq:Hred} are sparse $2066\times2066$ matrices on $\mathcal V$. Note that the generators $K_{S,ij}$, \textit{etc.}, have to be squared before reducing to $\mathcal V$. On the walls $b=0$ and $c=0$, the channels lie in the kernels of $K_{S,23}^2$ and $K_{S,12}^2$, respectively, and on their intersection $r_1=r_2=r_3$ in the joint kernel. There, the initial channels are the lowest eigenvectors of the averaged $A$ within the kernel. In each enrichment round, we compute at every node the residual vector $R_n$, with components $(R_n)_{\tilde s}=\langle\eta N_n\otimes\tilde s|H_{\rm red}-E_h|\Psi_h\rangle$, $\tilde s=1,\dots,2066$. By virtue of the eigenvalue equation, $R_n$ is orthogonal to the channels of the node. To improve the variational solution, at every node we add $R_n$ along with its Davidson correction to the channels. On the finest meshes we instead add earlier solutions, interpolated to the new nodes, as channels, next to the lowest local ones, and solve once. The best state, with five channels per node, is obtained in this way from the solution with elements twice as long along the valley.

In each round, we have to compute the matrix elements of $H_{\rm red}$, as well as the overlaps between the basis functions. Two basis functions overlap only if their nodes belong to a common element, which implies that the corresponding matrices are sparse. We find the lowest eigenstate by inverse iteration, \textit{i.e.} by applying $(H_{\rm red}+1)^{-1}$ repeatedly to a few trial states, which enhances the state of lowest energy. Since $H_{\rm red}\geq0$, the operator $H_{\rm red}+1$ is positive definite, and we invert it using a sparse Cholesky factorization. For our best state, with $4.4\times10^5$ unknowns, the residual of the eigenvalue equation falls below $10^{-8}$, so that the linear algebra manipulations contribute negligibly to the error.

\subsection{Asymptotic tail of the wavefunction}

Given the numerical solution for the ground state $\Psi_h$, we define the marginal density of the largest singular value,
\begin{equation}\label{eq:full_PR}
\mathcal P(R)=\int d^3rJ\|\Psi_h\|^2\delta(R-r_1) \ , \qquad \int_0^\infty dR\,\mathcal P(R)=1 \ .
\end{equation}
Along the valley, $\mathcal P(R)$ coincides with the density~\eqref{eq:PR} of the separation of the two D0-branes, but, unlike the latter, it remains well defined in the strongly coupled core of the state. In the valley, $\mathcal P(R)=\tfrac89\Om R^8f^2(R)$, where the $\ell=2$ coefficient $f(R)$ of the relative wavefunction obeys the free radial equation in nine dimensions,
\begin{equation}
-f''(R)-\frac8Rf'+\frac{18}{R^2}f=k^2f \ ,
\end{equation}
with $k^2$ the energy $E_h$ of the state. For $k=0$, its decaying solution is the tail $f=CR^{-9}$. The computed state, however, is variational and has a small positive energy $E_h$, so its far field solves the free equation at $k^2=E_h$, not at $k=0$. It is then convenient to work with the amplitude $u=\sqrt{\mathcal P}=(\tfrac89\Om)^{1/2}R^4f$, which obeys
\begin{equation}
-u''+\frac{30}{R^2}u=k^2u \ , \qquad k^2=E_h \ ,
\label{eq:radial_app}
\end{equation}
a Bessel equation, with general solution
\begin{equation}
u(R)=c_1\sqrt RJ_{-11/2}(kR)+c_2\sqrt RJ_{11/2}(kR) \ .
\label{eq:Bessel_soln}
\end{equation}
The first term is the tail at energy $E_h$. As $k\to0$, it becomes $c_1(k/2)^{-11/2}R^{-5}/\Gamma(-\tfrac92)$, so that its coefficient determines the tail coefficient through $\tfrac89\Om|C|^2=[c_1(k/2)^{-11/2}/\Gamma(-\tfrac92)]^2$. The second term, which grows as $R^6$ for $kR\ll1$, describes the reflection off the wall of the box.

Numerically, $\mathcal P(R)$ is evaluated with $8\times8$ Gauss points per element in $(a,b)$. We then extract $c_1$ from our best state, with $L_c=22$, quartic polynomials on the $10\times10$ mesh and $\tau=10^{-16}$, for which $E_h=0.0003$. In the valley, the solution~\eqref{eq:Bessel_soln} holds up to corrections of relative order $(r_\perp/R)^2\sim R^{-3}$, where $r_\perp\sim R^{-1/2}$ is the transverse size. Hence, we fit $\sqrt{\mathcal P}$ to
\begin{equation}
u_{\rm fit}=\sqrt R\Big[c_1J_{-11/2}(kR)\Big(1+\sum_{p\in\mathcal S}D_pR^{-p}\Big)+c_2J_{11/2}(kR)\Big] \ , \qquad k^2=E_h \ ,
\end{equation}
with $\mathcal S=\{3\}$ or $\{3,6\}$, minimizing the relative deviation $\sum_j[u_{\rm fit}(R_j)/\sqrt{\mathcal P(R_j)}-1]^2$ over points spaced by $0.05$. The fitting windows start at $R_{\min}=7,8,9,10$, beyond the core, and end at $R_{\max}=17,18,19$ (well before the wall). The resulting $24$ fits give $|C|^2=286.39$, with a range $286.36$--$286.40$. We then change one parameter at a time in the same box and repeat the fit. With elements along the valley of at most $0.5$ and $0.35$ the fit gives $286.54$ and $286.43$. The $12\times12$ mesh and lowering $\tau$ to $10^{-24}$ change the fit by less than $0.01$, both measured with elements of $0.5$ along the valley, which gives~\eqref{eq:Cdirect}. Its uncertainty is dominated by the elements along the valley, for which we conservatively take the full change from $0.5$ to $0.25$ ($0.15$), rounded up.

The $R^{-3}$ correction is accurately captured: the local coefficient, \textit{i.e.} the ratio of $\mathcal P$ to the density of a tail with $|C|^2=1$ at $k^2=E_h$, exceeds its asymptote by $1.9\%$, $0.6\%$ and $0.2\%$ at $R=7$, $10$ and $15$, and the single term $D_3=3.1$ reproduces these values to two parts in $10^4$.

\subsection{Moments}

The moments of Tables~\ref{tab:moments} and~\ref{tab:obs} are obtained as follows. In each state, a moment is integrated inside the box up to a matching radius $R_c=8$, $10$ or $12$, and it is completed beyond that radius with the analytic tail~\eqref{eq:tail_three_grav_coupling_prediction}, with the state renormalized by the completed probability. For an observable that grows as $R^e$ along the valley, with $e<9$, the completion contributes
\begin{equation}
\Delta\langle R^e\rangle\simeq\frac{7560}{9-e}R_c^{e-9} \ ,
\end{equation}
which at $R_c=10$ and for $e=8$ amounts to about $4.7\%$ of $\vev{t_1^4}$, even though less than $10^{-6}$ of the probability lies beyond $R_c$. We quote the values obtained from our best state. The error is obtained by adding in quadrature the following contributions: the transverse remainder, which is the slope in $E_h$ of the refinement from cubic to quartic elements times the $0.0001$ of $E_h$ that the valley mesh does not account for; the full shift measured when the elements along the valley are halved; twice the change in one more enrichment round; the change with the pruning tolerance; the range over the three matching radii; and, for $\vev{V_B}$, its difference from $\vev{T}$.

Note that the moments obtained here refer to the BFSS Hamiltonian~\eqref{eq:H} with $g=2$. The literature on the matrix bootstrap~\cite{Lin23,LinZheng}, instead, typically works with the Hamiltonian,
\begin{equation}
H=\Tr\Big(\tfrac12g^2\big(P^I\big)^2-\tfrac{1}{4g^2}[X^I,X^J]^2-\tfrac12\psi^T\gamma^I[X^I,\psi]\Big) \ ,
\end{equation}
in units where $g=1$. For $SU(2)$ we use the unnormalized trace. The two Hamiltonians are related by $X\to gX$, $P\to P/g$, combined with the scaling $H(g)=g^{2/3}h$, so that
\begin{equation}
\langle\mathcal O(X,P,\psi)\rangle_\text{here}=2^{\frac{n_P-n_X}{3}}\langle\mathcal O(X,P,\psi)\rangle_\text{MB} \ ,
\end{equation}
where $n_X$ and $n_P$ are the numbers of $X$ and $P$ letters in the word $\mathcal O$. We are interested in the moments $\vev{\Tr X^{2n}}\equiv\vev{\Tr(X^IX^I)^n}$ and $\vev K=\tfrac12\vev{\Tr P^IP^I}$. For $SU(2)$, $X^IX^I=\tfrac14t_1\mathds{1}$ with $t_1=\sum_{I,A}q_{IA}^2=r_1^2+r_2^2+r_3^2$, and $T=\sum_{I,A}p_{IA}^2$ is the kinetic energy, so that
\begin{equation}
\vev{\Tr X^{2n}}_\text{MB}=2^{1-4n/3}\vev{t_1^n}_\text{here} \ , \qquad \vev K_\text{MB}=2^{-2/3}\vev T_\text{here} \ .
\end{equation}
Table~\ref{tab:obs} contains the moments in both normalizations, with $\vev K$ obtained from $\vev{V_B}$ through the virial relation $\vev T=\vev{V_B}$. These moments are finite only for $n\le4$, because of the $R^{-10}$ tail of the density.

\par\noindent\begin{minipage}{\linewidth}
\begin{center}
\small\setlength{\tabcolsep}{8pt}
\begin{tabular}{@{}lrlr@{}}
\toprule
moment & $g=2$ & moment & $g=1$\\
\midrule
$\vev{t_1}$ & $10.1729(7)$ & $\vev{\Tr X^2}$ & $8.0742(6)$\\
$\vev{t_1^2}$ & $109.14(2)$ & $\vev{\Tr X^4}$ & $34.378(6)$\\
$\vev{t_1^3}$ & $1244.9(4)$ & $\vev{\Tr X^6}$ & $155.61(5)$\\
$\vev{t_1^4}$ & $16140(10)$ & $\vev{\Tr X^8}$ & $800.6(5)$\\
$\vev{V_B}=\tfrac12\vev{t_1^2-t_2}$ & $27.164(3)$ & $\vev K$ & $17.112(2)$\\
$\vev{t_2}$ & $54.82(2)$ & & \\
$\vev{t_3}$ & $407.2(3)$ & & \\
\bottomrule
\end{tabular}
\end{center}
{\small\refstepcounter{table}\label{tab:obs}\noindent TABLE~\thetable.\ Moments of the bound state. Left: in our units, $g=2$, with $t_n=\sum_ir_i^{2n}$. Right: in the normalization of the matrix bootstrap, $g=1$.\par}
\end{minipage}
\medskip

\section{Details on the Calculation of the Phase Shifts}\label{app:phase}

In this Appendix we spell out the computation of Section~\ref{sec:ward}. The interaction $K(k_f,k_i)$ between two gravitons, given in eq.~\eqref{eq:bornkernel} below, is invariant under simultaneous rotations of the relative momenta and of the polarizations, and therefore it preserves the total $Spin(9)$ representation. Consequently, in a sector that contains a single two-graviton channel, such as the singlet and the $\rep{495}$ we are considering, the interaction cannot mix this channel with any other, and the channel scatters with a single phase shift. In the first Born approximation~\cite{LandauLifshitz}, the phase shift of a channel is given by
\begin{equation}
\tan\delta=-\frac{\pi k^7}{2(2\pi)^9}\int d\Omega_fd\Omega_i\Psi^\dagger(\hat k_f)K(k_f,k_i)\Psi(\hat k_i) \ ,
\label{eq:born}
\end{equation}
where $k_i=k\hat k_i,\,k_f=k\hat k_f$ are the relative momenta before and after the scattering, $K(k_f,k_i)$ is the interaction between them and $\Psi(\hat k)$ is the (normalized) angular wavefunction of the channel.

\textit{The long-range interaction.} At long distances, the interaction of two D0-branes is~\cite{PSW98,NicolaiPlefka}
\begin{equation}
K=-\frac{\pi^4}{4q^2}\Big[(v^2)^2-v^2A+\frac{A^2}2-\frac{AB+BA}{36}+\frac{B^2}{504}\Big] \ ,
\label{eq:bornkernel}
\end{equation}
where $q=k_f-k_i$ is the momentum transfer, $v=k_i+k_f$ the relative velocity in our units, and
\begin{equation}
A=v_Iq_J\mathcal J^{IJ} \ , \qquad B=\sum_M(q_I\mathcal J^{IM})(q_J\mathcal J^{JM}) \ , \qquad \mathcal J^{IJ}=\theta_\alpha\gamma^{IJ}_{\alpha\beta}\theta_\beta \ ,
\end{equation}
where $\mathcal J^{IJ}$ generate the rotations of the polarizations, so that on the $\rep{44}$ they act as on a symmetric tensor, \textit{i.e.}
\begin{equation}
\mathcal J^{IJ}\ket{h_{KL}}=4\big(\delta_{JK}\ket{h_{IL}}-\delta_{IK}\ket{h_{JL}}+\delta_{JL}\ket{h_{KI}}-\delta_{IL}\ket{h_{KJ}}\big) \ .
\label{eq:Jon44}
\end{equation}
For a scattering angle $\chi$, we also have
\begin{equation}
v^2=2k^2(1+\cos\chi) \ , \qquad q^2=2k^2(1-\cos\chi) \ .
\end{equation}
Note that eq.~\eqref{eq:bornkernel} holds up to contact terms~\cite{PSW98}, namely terms without the $1/q^2$ pole. By dimensional analysis these are polynomials of degree two in the momenta, whereas the projection onto a $d$-wave in eq.~\eqref{eq:born} requires a degree of at least two in each of $k_i$ and $k_f$, so they do not contribute.

\textit{The two channels.} We are interested in the singlet and in the $\rep{495}$, both of which have polarization $\rep{44}$. The normalized wavefunction of the singlet harmonic on the eight-sphere reads,
\begin{equation}
\Psi_{\rep1}(\hat k)=\Big(\frac{9}{8\,\Om}\Big)^{1/2}\,\hat k_I\hat k_J\ket{h_{IJ}} \ ,
\label{eq:psi1}
\end{equation}
while the one of the $\rep{495}$, up to a normalization constant,
\begin{equation}
\Psi_{\rep{495}}(\hat k)\propto R_{IKJL}\,\hat k_K\hat k_L\ket{h_{IJ}} \ ,
\end{equation}
where $R_{IKJL}$ is traceless and has the symmetries of the Riemann tensor. Using eq.~\eqref{eq:Jon44} and the normalization of $|h_{IJ}\rangle$ given after eq.~\eqref{eq:tail}, we can now evaluate eq.~\eqref{eq:born} for the two channels. Doing so, we find the following phase shifts
\begin{align}
\tan\delta_{\rep1}\big|_{K}&=\frac{k^9}{8!}\Big(1+12+84+\frac{448}3+96\Big)=\frac{k^9}{8!}+\frac8{945}k^9 \ ,\\
\tan\delta_{\rep{495}}\big|_{K}&=\frac{k^9}{8!}\Big(1+\frac43+\frac45-\frac{64}{15}+\frac{32}{15}\Big)=\frac{k^9}{8!} \ .
\end{align}
The five terms in each sum are those of eq.~\eqref{eq:bornkernel}, in the same order.

\textit{The bound state.} In the singlet there is one more contribution at order $k^9$. The exact scattering amplitude between $k_i$ and $k_f$ is the matrix element of
\begin{equation}
T=V+V\frac{1}{E-H}V \ ,
\end{equation}
where $H=k^2+V$ is the relative Hamiltonian and $E=k^2$ is the energy. The first term is the interaction itself, which at long distances is $K$. In the second term the two gravitons interact, propagate, and interact again. The bound state appears in the propagator as a pole at zero energy,
\begin{equation}
\frac{1}{E-H}=\frac{\ket{B}\bra{B}}{E}+\dots \ ,
\end{equation}
where $\langle B|B\rangle=1$ and the dots stand for the states of the continuum. Two gravitons can thus merge into the bound state and split again, which adds to the interaction the term
\begin{equation}
K_B(k_f,k_i)=\frac{1}{k^2}\bra{k_f}V\ket{B}\bra{B}V\ket{k_i} \ , \qquad \bra{k}V\ket{B}=C\frac{\Om}{9}k_Ik_J\ket{h_{IJ}} \ .
\end{equation}
The second expression is the three-point vertex of eq.~\eqref{eq:split-vertex}. With the wavefunction of eq.~\eqref{eq:psi1}, it yields
\begin{equation}
\int d\Omega\,\Psi_{\rep1}^\dagger(\hat k)\,\bra{k}V\ket{B}=C\,\frac{\Om}{9}\Big(\frac{8\,\Om}{9}\Big)^{1/2}k^2 \ ,
\end{equation}
and eq.~\eqref{eq:born}, with $K_B$ in the place of $K$, gives
\begin{equation}
\tan\delta_{\rep1}\big|_{B}=-\frac{\pi k^7}{2(2\pi)^9}\frac{8\Om^3}{729}|C|^2k^2=-\frac{256\pi^4}{945^3}|C|^2k^9 \ .
\label{eq:boundterm}
\end{equation}
Importantly, the remaining contributions do not enter at order $k^9$. The iterations of the long-range interaction and the higher-loop corrections, which by the non-renormalization theorems~\cite{PSS} start at order $v^6$, carry additional powers of $k$. As in Section~\ref{sec:ward}, we assume that the short-distance region, where the matrices do not commute, acts as an interaction of short range and hence is also suppressed at this order.

\textit{The tail coefficient.} At order $k^9$, the two phase shifts therefore read,
\begin{equation}
\tan\delta_{\rep1}\simeq\frac{k^9}{8!}+\frac8{945}\,k^9-\frac{256\,\pi^4}{945^3}\,|C|^2k^9 \ , \qquad \tan\delta_{\rep{495}}\simeq\frac{k^9}{8!} \ .
\end{equation}
By the super-Ward identity, eq.~\eqref{eq:equalphases}, the phase shifts have to coincide. That happens provided the three-point coupling takes the value of eq.~\eqref{eq:tail_three_grav_coupling_prediction}.